\documentclass[onecolumn,iop,preprint]{emulateapj}

\usepackage{graphicx}
\usepackage{subfigure}
\usepackage{color}
\usepackage{amssymb}
\usepackage{amsmath}
\usepackage{enumerate}

          \font\sixrm=cmr6       

\def\teq#1{$\, #1\,$}                         

\def\dover#1#2{\hbox{${{\displaystyle#1 \vphantom{(} }\over{
\displaystyle #2 \vphantom{(} }}$}}
{\catcode`\@=11                                                  
\gdef\SchlangeUnter#1#2{\lower2pt\vbox{\baselineskip 0pt\lineskip0pt    
\ialign{$\m@th#1\hfil##\hfil$\crcr#2\crcr\sim\crcr}}}}           
\def\gtrsim{\mathrel{\mathpalette\SchlangeUnter>}}               
\def\lesssim{\mathrel{\mathpalette\SchlangeUnter<}}

\def\gamth{\gamma_{\hbox{\sixrm th}}}
\def\gammaMin{\gamma_{\hbox{\sixrm min}}}

\def\vFv{\nu F_{\nu}}

\newcommand{\beq}{\begin{equation}}
\newcommand{\eeq}{\end{equation}}
\newcommand{\g}{\gamma}

\def\e{\epsilon}
\def\ep{\epsilon^\prime}

\newcommand{\vol}[2]{$\,$\rm #1\rm , #2.}

\shorttitle{Time-resolved Analysis of Fermi GRBs with Physical Synchrotron
  Photon Models}
\shortauthors{J.~Michael Burgess et al.}

\begin{document}

\title{Time-resolved Analysis of Fermi GRBs with Fast and Slow-Cooled Synchrotron Photon Models}

\author{J.~M.~Burgess\altaffilmark{1,2},
R.~D.~Preece\altaffilmark{1,3},
V.~Connaughton,\altaffilmark{1},
M.~S.~Briggs\altaffilmark{1},
A.~Goldstein\altaffilmark{1},
P.~N. Bhat\altaffilmark{1},
J.~Greiner\altaffilmark{4},
D.~Gruber\altaffilmark{4}, 
A.~Kienlin\altaffilmark{4},
C.~Kouveliotou\altaffilmark{5}, 
S.~McGlynn\altaffilmark{6},
C.~A.~Meegan\altaffilmark{7}, 
W.~S.~Paciesas\altaffilmark{1},
A.~Rau\altaffilmark{4},
S.~Xiong\altaffilmark{1}, 
}

\author{M.~Axelsson\altaffilmark{8,9,10}, 
M.~G.~Baring\altaffilmark{11,12}, 
C.~D.~Dermer\altaffilmark{13}, 
S.~Iyyani\altaffilmark{9,14},
D.~Kocevski\altaffilmark{15}, 
N.~Omodei\altaffilmark{15}, 
F.~Ryde\altaffilmark{10,9}, 
G.~Vianello\altaffilmark{15,16}
}

\altaffiltext{1}{University of Alabama in Huntsville, 320 Sparkman
  Drive, Huntsville, AL 35899, USA}


\altaffiltext{2}{email: james.m.burgess@nasa.gov}
\altaffiltext{3}{email: Rob.Preece@nasa.gov}

\altaffiltext{4}{Max-Planck-Institut f$\rm \ddot{u}$r
  extraterrestrische Physik (Giessenbachstrasse 1, 85748 Garching,
  Germany)}


\altaffiltext{5}{Space Science Office, VP62, NASA/Marshall
  Space Flight Center, Huntsville, AL 35812, USA}

\altaffiltext{6}{Exzellence Cluster "Universe", Technische Universität
  München, Boltzmannstrasse 2, 85748, Garching, Germany}

\altaffiltext{7}{Universities Space Research Association, 320 Sparkman
  Drive, Huntsville, AL 35899, USA}

\altaffiltext{8}{Department of Astronomy, Stockholm University, SE-106
  91 Stockholm, Sweden}

\altaffiltext{9}{The Oskar Klein Centre for Cosmoparticle Physics,
  AlbaNova, SE-106 91 Stockholm, Sweden}

\altaffiltext{10}{Department of Physics, Royal Institute of Technology
  (KTH), AlbaNova, SE-106 91 Stockholm, Sweden}

\altaffiltext{11}{Rice University, Department of Physics and
  Astronomy, MS-108, P. O. Box 1892, Houston, TX 77251, USA}

\altaffiltext{12}{email: baring@rice.edu}

\altaffiltext{13}{Space Science Division, Naval Research Laboratory,
  Washington, DC 20375-5352, USA}

\altaffiltext{14}{email: shabuiyyani@gmail.com}

\altaffiltext{15}{W. W. Hansen Experimental Physics Laboratory, Kavli
  Institute for Particle Astrophysics and Cosmology, Department of
  Physics and SLAC National Accelerator Laboratory, Stanford
  University, Stanford, CA 94305, USA}

\altaffiltext{16}{Consorzio Interuniversitario per la Fisica Spaziale
  (CIFS), I-10133 Torino, Italy}

\email{james.m.burgess@nasa.gov}

\begin{abstract}
 
  Time-resolved spectroscopy is performed on eight bright, long
  gamma-ray bursts (GRBs) dominated by single emission pulses that
  were observed with the {\it Fermi Gamma-ray Space
    Telescope}. Fitting the prompt radiation of GRBs by empirical
  spectral forms such as the Band function leads to ambiguous
  conclusions about the physical model for the prompt
  radiation. Moreover, the Band function is often inadequate to fit
  the data. The GRB spectrum is therefore modeled with two emission
  components consisting of optically thin non-thermal synchrotron
  radiation from relativistic electrons and, when significant, thermal
  emission from a jet photosphere, which is represented by a blackbody
  spectrum. To produce an acceptable fit, the addition of a blackbody
  component is required in 5 out of the 8 cases. We also find that the
  low-energy spectral index $\alpha$ is consistent with a synchrotron
  component with $\alpha = -0.81\pm 0.1$. This value lies between the
  limiting values of $\alpha = -2/3$ and $\alpha = -3/2$ for electrons
  in the slow and fast-cooling regimes, respectively, suggesting
  ongoing acceleration at the emission site. The blackbody component
  can be more significant when using a physical synchrotron model
  instead of the Band function, illustrating that the Band function
  does not serve as a good proxy for a non-thermal synchrotron
  emission component. The temperature and characteristic
  emission-region size of the blackbody component are found to,
  respectively, decrease and increase as power laws with time during
  the prompt phase. In addition, we find that the blackbody and
  non-thermal components have separate temporal behaviors as far as
  their respective flux and spectral evolutions.

\end{abstract}



\section{Introduction}

\label{sec:intro}
The fireball model of GRBs
\citep{Cavallo:1978,Goodman:1986,Paczynski:1986} assumes that a large
amount of energy is released in a small space leading to a fireball
that emits $\gamma$-rays. The progenitors of these events are not
known, but are believed to be the collapse of massive stars, or the
coalescence of two compact objects. Details of this model rely heavily
on assumptions about the initial parameters of the fireball including
the radii of emission, the baryon load and the magnetic field
associated with the plasma outflow (for review see
\citet{Meszaros:2006}). The observed spectra of GRBs are typically
interpreted as non-thermal \citep{Piran:1999,Zhang:2012} though there
are some exceptions \citep{Ryde:2004}. This interpretation indicates
that a dissipative process such as internal shocks energizes the
electrons to non-thermal distributions. Though some versions of the
fireball model predict that the entire spectrum in the $\gamma$-ray
band be in the form of thermal emission, we concentrate on the version
presented in \citet{Meszaros:2000} where a mixture of thermal and
non-thermal (synchrotron) emission emerges. Alternative models such as
the Poynting-flux dominated outflow (PFD) model exist and could
provide a viable mechanism for generating the observed prompt emission
\citep[e.g.,][]{2009JPhCS.189a2018G,Zhang:2011}. These PFD models have
not yet advanced to quantitative spectral and temporal predictions;
therefore, it is difficult to compare our results to these
models as extensively as is possible for the internal shock model for
which extensive simulations of lightcurves and spectra have been
developed \citep[e..g.,][]{Daigne:1998,Daigne:2009}. The physical
emission process behind these extreme cosmic explosions has not been
established. In the past, GRB spectra have been well fit by the Band
function \citep{Band:1993p20784}, which is an exponentially cutoff
power law that smoothly joins to a second power law, and is
parameterized by the function

\begin{equation}
\label{eq:band}
 F_{\nu}(\mathcal{E})\;=\;F_0
  \begin{cases}
    {\left(\frac{\mathcal{E}}{\rm 100\;keV}\right)^{\alpha} {\rm exp}\left(-\frac{(2+\alpha)\mathcal{E}}{E_{\rm p}}\right),} & {\mathcal{E}\le (\alpha-\beta)\frac{E_{\rm p}}{ (\alpha+2)} }\\
    {\left(\frac{\mathcal{E}}{\rm 100\;keV}\right)^{\beta}{\rm
        exp}{\left(\beta-\alpha\right)}\left[\frac{(\alpha-\beta)E_{\rm
            p}}{{\rm 100\;keV} (2+\alpha)}\right]^{\alpha-\beta}}, &
    {\mathcal{E}>(\alpha-\beta)\frac{E_{\rm p}}{ (\alpha+2)}\;, }
  \end{cases}
\end{equation}
where $\mathcal{E}$ is the photon energy. The Band function is
characterized by low and high-energy power-law spectral indices (the
Band $\alpha$ and Band $\beta$ parameters, respectively) as well as
its $\vFv$ peak energy, E$_{\rm p}$. The ability of the Band function
to fit most prompt-spectra has led to the widespread use of the Band
function's fitted parameters as indicators of the underlying emission
and acceleration processes. In particular, the value of $\alpha$ has
been of interest because non-thermal spectra and indices from
synchrotron, inverse Compton, and other processes are each constrained
by allowed indices at energies less than their respective $\vFv$
peaks. Above the peak, the high-energy index is variable and related
to the distribution of the emitting electrons and provides fewer clues
for discerning between emission processes. Although such
extrapolations can be useful, \citet{Burgess:2011} (hereafter B11)
directly fit a synchrotron photon model to the prompt emission data of
GRB 090820A, showing that it is possible to fit a physical model
directly to the data without relying on interpretations of the Band
function in order to determine the $\gamma$-ray emission process.

Previous studies
\citep{Gonzalez:2003,Ryde:2005,Guiriec:2010,Ryde:2010,Guiriec:2013}
showed that a single Band function cannot fully account for the
spectrum of all GRBs. The addition of a blackbody and/or power-law
better describes the spectra of these GRBs. Specifically, the addition
of the blackbody below E$_{\rm p}$ also allows for the direct fitting
of the synchrotron model to the data, which we exploit in this work.
The addition of a blackbody in the spectrum allows a quantitative
analysis of its properties to ascertain its origin. Several studies
\citep{Meszaros:2000,Daigne:2002,Ryde:2004,Ryde:2006} have developed
the theoretical framework for a blackbody component coexisting with a
non-thermal component in the spectra of GRBs, as well as testable
relations that can be applied to data to investigate the photosphere
of GRBs.

Herein, we extend the analysis of B11 to several bright {\it Fermi}
GRBs and in addition investigate the ability of the Band function to
serve as a proxy for physical emissivities in the fitting process. We
show that slow-cooled synchrotron is a viable model for GRB prompt
emission using data from both the Gamma-ray Burst Monitor (GBM) (10
keV~-~40 MeV) and Large Area Telescope (LAT) (100 MeV~-~300 GeV) on
board the {\it Fermi} space telescope. This analysis makes use of the
newly released LAT low-energy (LLE) data which provides a larger
effective area between 30 MeV and $\sim$1 GeV, allowing us to extend
the LAT analysis down to 30 MeV \citep[see][and work in
preparation]{Pelassa:2010}. First, we give a description of the
non-thermal and blackbody photon emissivities in Section
\ref{sec:model}. We then present our analysis technique in Section
\ref{sec:observe} and compare the results with predictions about each
component in Section \ref{sec:results}.  We discuss the results in
Section \ref{sec:discussion}, and constraints on synchrotron shell
models are presented in the Appendix.


\section{Model Spectral Components}
\label{sec:model}
In the fireball model of GRBs, the majority of the flux is
theoretically expected to be in the form of thermal emission coming
from the photosphere of the jet. However, nearly all of the low-energy
indices implied by GRB spectral analysis with Band-function spectral
inputs have $\alpha<+1$, i.e. too soft to be thermal -- see for
example \cite{Gold:2012a,Gold:2012b} for the BATSE database.  This
points to a non-thermal emission process for most GRBs. Multi-spectral
component analysis of {\it Fermi} GRBs has shown that while the
majority of the emission is non-thermal, a small fraction of the
energy radiated apparently originates from a blackbody component
\citep{Guiriec:2011,Axelsson:2012}. In B11, a blackbody component was
also identified when the non-thermal emission was fit with a
synchrotron photon model. This combination of blackbody and
non-thermal emission was predicted by \citet{Meszaros:2000}.  In this
paper, before proceeding to the details of the analysis, we first
review the synchrotron model (see B11), and also several observable
relations of the blackbody component. These components are then
implemented into a fitting program which directly convolves the
physical models with the GBM detector response to compare with
observations.

\subsection{Synchrotron Radiation}
\label{sec:model:non-thermal}
While no definitive model for the non-thermal emission of GRBs exists,
electron synchrotron is the simplest and most efficient non-thermal
emission process that could account for the prompt GRB radiation. It
is a natural first choice in moving from the traditional use of
empirical fitting functions to physical models, which will then afford
more reliable tests for characterizing the GRB environment.  In B11,
we implemented a parameterized synchrotron model for fitting GRB data
directly. We implemented a synchrotron model first used in fitting the
deconvolved spectra of GRBs by \citet{Tavani:1996p23809} and later by
\citet{Baring:2004p10999}. \citet{Sari:1998} identified two regimes
for the synchrotron emission which are relevant for GRBs: fast and
slow-cooling. The difference between the two is related to the
radiative time-scale of the emission.

\begin{figure}

  \centering

  \includegraphics[scale=1, angle=0]{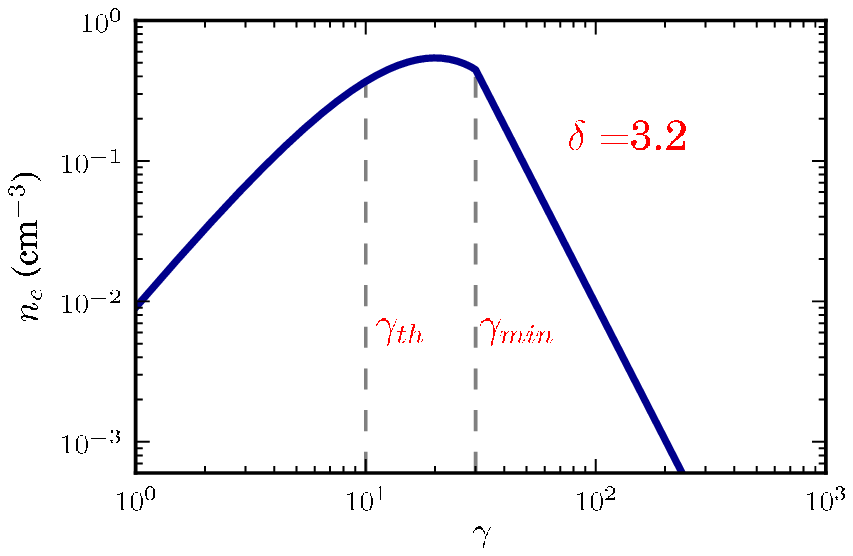}

  \caption{The physical electron spectrum (see eq. \ref{eq:elec_dist})
    used for fitting synchrotron spectra directly to the data. Here
    $\gamth=300$ and $\gammaMin=900$ which are the fixed values used
    for all fits in this study. The value of $\epsilon$ is
      fixed to $(\gammaMin/\gamth)^2\times exp(-\gammaMin/\gamth)$.}

  \label{fig:slowE}

\end{figure}

For the slow-cooling synchrotron model, we assume the electron
distribution from B11 (see also \citet{Baring:2004p10999}) which
includes electrons in a thermal pool as well as electrons that are
accelerated into a power-law tail (see Figure \ref{fig:slowE}), given by
\begin{equation}
  n_e(\gamma )\; =\; n_{0} \biggl\lbrack\;
  \Bigl( \dover{\gamma}{\gamth} \Bigr)^2\,
  e^{-\gamma/\gamth } + \epsilon \,
  \Bigl( \dover{\gamma}{\gamth} \Bigr)^{-\delta}\,
  \Theta \Bigl( \dover{\gamma}{\gammaMin} \Bigr)\, \biggr\rbrack\, .
  \label{eq:elec_dist}
\end{equation}
Here, $n_0$ normalizes the distribution to total number or energy,
$\gamma$ is the electron Lorentz factor in the fluid frame, $\gamth$
is the thermal electron Lorentz factor, $\gammaMin$ is the minimum
electron Lorentz factor of the power-law tail, $\epsilon$ is the
normalization of the power-law, and $\delta$ is the electron spectral
index. The function $\Theta(x)$ is a step function where $\Theta(x)=0$
for $x<1$ and $\Theta(x)=1$ for $x>1$.  In analyses of the shock
acceleration process, distributions like eq.\ (\ref{eq:elec_dist}) are
realized for $\epsilon\ll 1$ regimes both for non-relativistic shocks
\citep[e.g.,][]{Baring:1995}, and also ones where the upstream flow
impacts the shock layer relativistically
\citep{Ellison:1990,Ellison:2004,Spitkovsky:2008,Baring:2011,Baring:2012}.
In all these works, the charges (in this case electrons) are
accelerated directly from the thermal pool to generate a power-law
distribution corresponding to $\epsilon\lesssim 0.1$ that smoothly
connects to the thermal distribution.  This is no surprise since
strong dissipation is present in plasma shocks, and so thermalization
is prominent for many (but not all) particles.  These results motivate
our general form for the electron distribution, noting that our
parametrization may oversimplify the process of acceleration and
ignores the effects of photon-electron collisions such as Compton
scattering that are likely taking place in the high $\gamma$-ray
regime. In particular, as pointed out by \citep{Baring:2004p10999},
the distribution in Eq.~(\ref{eq:elec_dist}) is only viable for
approximating GRB spectra with synchrotron emission when $\epsilon\gg
1$, and the power-law component dominates the Maxwell-Boltzmann one.
This constraint is accommodated in our fitting protocol, and we
discuss its value shortly.

Finally, note that simulations that include the full radiative heating
and cooling of electrons expected in the high $\gamma$-ray regime but
that neglect acceleration processes produce electron distributions
that differ greatly from simple power-laws \citep{Peer:2004}. These
complicated electron distributions are difficult to model in a fitting
process and may require spectral resolution beyond what the {\it
  Fermi} data provide to identify in spectral fits.






We convolve this simplified distribution with the
standard isotropic synchrotron kernel \citep{rybicki}
\begin{equation}
  F_{\nu}(\mathcal{E})\; \propto\; \int_1^{\infty} n_e(\gamma ) \, 
  \mathcal{F} \left( \dover{\mathcal{E}}{\mathcal{E}_c}\right) \, d\gamma\quad ,
  \label{eq:synch_flux}
\end{equation}
where%
\begin{equation}
  \mathcal{F}\left(w\right) \; =\; w \int_w^{\infty } K_{5/3}(x) \, dx
  \label{eq:synch_func}
\end{equation}
expresses the single-particle synchrotron emissivity (i.e., energy per
unit time per unit volume) in dimensionless functional form. The
$K_{5/3}$ term is the modified Bessel function of the second kind. The
characteristic scale for the synchrotron photon energy is
\begin{equation}
  \mathcal{E}_c\; =\; E_* \gamma^2\ ,
  \label{eq:ergcsynch}
\end{equation}
where $E_*\equiv 3\Gamma B /(2B_{cr})\, m_{e}c^2$, B is the
magnetic field strength, $\Gamma$ is the bulk Lorentz factor,
and \teq{B_{\rm cr} = 4.41 \times 10^{13}}Gauss is the quantum
critical field.

In principle, there are six spectral parameters that can be
constrained by the fits: $n_0$, $E_{*}$, $\delta$, $\epsilon$,
$\gamth$, and $\gammaMin$; however, we fix $\gamth$, $\gammaMin$, and
$\epsilon$ due to fitting correlations as explained below. The
parameter $E_{*}$ scales the energy of the fit and is linearly related
to the Band function's E$_{\rm p}$. Numerical simulations of particle
acceleration at relativistic shocks have shown \citep[see references
in][]{Baring:2004p10999} that the non-thermal population is generated
directly from the thermal one.  To match these circumstances, we set
the ratio of $\gamth$ and $\gammaMin$ to be $\sim$3, following
\citep{Baring:2004p10999}.  The parameters $E_*$, $\gamth$, and
$\gammaMin$ all directly scale the peak energy of the spectrum but do
not alter its shape and thus cannot be independently determined. For
this reason we chose values of $\gamth=300$ and $\gammaMin=900$ for
all fits and left $E_*$ free to be constrained from the fit.

As shown in the Appendix, such parameter values are on the outer edge
for what is allowed energetically.  For these parameters, the flow is
strongly Poynting flux/magnetic-field energy dominated in order that
electrons with $\gamma\sim 300$ -- 900 can produce radiation in the
MeV regime.  Magnetized jet models are advantageous for energy
dissipation through magnetic reconnection, to produce short timescale
variability, and to accelerate ultra-high energy cosmic ray (see
Appendix). In fact, a wide range of parameter values with much larger
electron Lorentz factors and smaller magnetic fields are possible. In
weak magnetic-field models, a strong self-Compton component and
$\gamma\gamma$ opacity effects can make a cascade that modifies the
standard emission spectrum of GRBs. By considering a strongly
magnetically dominated model, these issues can be neglected.

For the chosen parameters, the system is always in the strongly cooled
regime (see Appendix). Nevertheless, we adopt the expression, eq.\
(\ref{eq:elec_dist}), to approximate an electron spectrum in the
slow-cooling regime.  The parameter $\epsilon$, corresponding to the
relative amplitude between the thermal and non-thermal portions of the
electron distribution, was not easily constrained in the fitting
process used in B11 and produced small non-physical discontinuities in
the electron distribution, as pointed out in
\citet{Beloborodov:2012}. Therefore, here we numerically fix this
parameter to the small value of $(\gammaMin/\gamth)^2\times
\exp(-\gammaMin/\gamth)$, so that there is no discernible
discontinuity between the thermal and non-thermal parts of the
distribution. The thermal component helps smooth out the
  spectral structure at $E_{\ast}$, but does not alter the asymptotic
  index of $\alpha = 2/3$ realized for synchrotron emission from
  populations with lower bounds to their particle energies. After
  these simplifications, three shape parameters remain free: $E_{*}$,
  $\delta$ and $n_0$, which corresponds to the amplitude. Compared
  with the Band function's four fit parameters this model is simpler
  yet tied to actual physical processes.

\begin{figure}

  \centering

  \includegraphics[scale=1, angle=0]{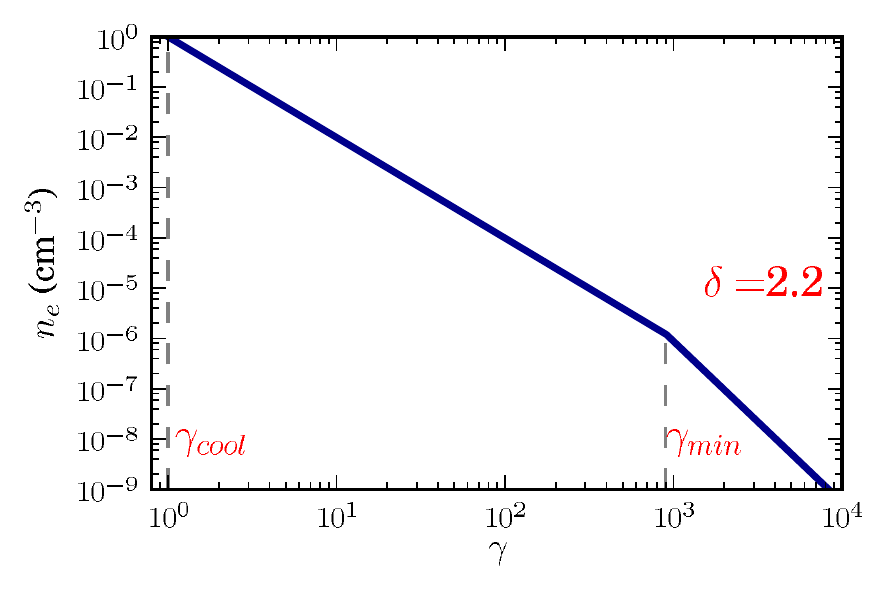}

  \caption{The fast-cooled electron distribution with
    $\gamma_{cool}=1$ and $\gammaMin=900$. A low-energy thermal
    distribution is excluded in this distribution because the
    synchrotron emission from the power-law form is already too broad
    for the typical $\vFv$ peak of the data and including a thermal
    distribution would smooth it further.}

  \label{fig:fastE}

\end{figure}

The second regime of synchrotron emission is the so-called
fast-cooling regime, which applies for the adopted parameters in a
naive cooling model. We assume that some acceleration process injects
a power-law distribution of electrons
\begin{equation}
  \label{eq:injected}
  N^{\rm inj}_e(\gamma)\;=\;n_e(\delta-1)\gammaMin^{(\delta-1)}\gamma^{-\delta}\;,\;\gammaMin\leq\gamma
\end{equation}
into a region where they are allowed to cool. Here, $n_e$ is the
electron density. We neglect the inclusion of the thermal pool because
its association with the radiative region is poorly
understood. Moreover, we show below that the fast-cooling spectrum is
already too broad for the typical GRB $\vFv$ peak and the thermal
distribution only broadens the spectrum further. Neglecting
  adiabatic losses and other acceleration mechanisms, the cooling of
electrons is governed by the continuity equation
\citep{Blumenthal:1970}
\begin{equation}
  \label{eq:komp}
  \frac{\partial n_e(\gamma,t)}{\partial t}+\frac{\partial}{\partial \gamma}[\dot{\gamma}n_e(\gamma,t)]+\frac{n_e(\gamma,t)}{t_{esc}}\;=\;N^{\rm inj}_e(\gamma)
\end{equation}
where $n_e/t_{esc}$ represents the loss of particles from the emission
region from which we can define the maximal cooling scale
$\gamma/\dot{\gamma}\;\sim\;t_{esc}$. This corresponds to a Lorentz
factor, $\gamma_{cool}$, below which cooling shuts off. We expect that
$\gamma_{cool}$ lies well below $\gammaMin$ owing to the fact that a
cooling break has not been observed in GRB data 
\citep[see, however,][]{Kocevski:2012}. For very high electron Lorentz factors,
$\gamma/\dot{\gamma}\;\ll\;t_{esc}$ and therefore the loss term can be
neglected, i.e., the dynamical timescale is much longer than the
radiative timescale. With this assumption we can simplify
eq. (\ref{eq:komp}) and it becomes time independent. The resulting
electron distribution can then easily be solved for
\begin{equation}
  \label{eq:simpKomp}
  n_e(\gamma,t)\;\approx\;\frac{1}{\dot{\gamma}}\int_{\gamma}^{\infty}N^{\rm inj}_e(\gamma')d\gamma'.
\end{equation}
Substituting in the synchrotron cooling rate
\begin{equation}
  \label{synchCool}
  \dot{\gamma}\;=\;-\frac{4}{9}\frac{r_0c}{r_g^2}\gamma^2\;,
\end{equation}
where $r_g\;=\;m_ec^2/eB$ and $r_0\;=\;e^2/m_ec^2$, eq. (\ref{eq:simpKomp}) yields the
synchrotron-cooled broken power-law distribution of electrons (see Figure \ref{fig:fastE})
\begin{equation}
  \label{eq:necool}
  n_e^{cool}\;\propto\;\frac{n_e\gammaMin}{\gamma^2}\;\min\left\{\left(\frac{\gamma}{\gammaMin}\right)^{-(\delta-1)},1\right\}\;,\;\gamma_{cool}\leq\gamma.
\end{equation}
This distribution is convolved with eq. (\ref{eq:synch_func}) to
produce a photon spectrum.  The radiation spectrum generated by this
distribution has an asymptotic low-energy index of $-$3/2, the
so-called ``second line-of-death" \citep{Cohen:1997}.  The free
parameters used for fitting this spectrum are the electron spectral
index, $\delta$, the fast-cooling equivalent of $E_*$ and the overall
amplitude.  However, the majority of GRB spectra in the BATSE
\citep{Gold:2012b} and GBM \citep{Gold:2012a} spectral catalogs have
spectra with low-energy indices harder than $-$3/2. Even if the
spectra were consistent with the low-energy index, the fast-cooling
spectrum's curvature around the $\vFv$ peak is much broader than
that observed around the peak of the spectrum (See
\ref{sec:fast}).

Even though we examine a model with $\gamma_{th} = 300$ and
$\gamma_{min} = 900$, the spectral fitting is insensitive to the exact
value of the product $\Gamma B\gamma^2$ provided that the constraints
discussed in Appendix A are satisfied. Even in a strong cooling regime
defined by these low assumed values of $\gamma_{th}$ and $\gamma_{min}
$, second-order processes in GRBs \citep{Waxman:1995,Dermer:2001},
which can become more important than first-order processes in
relativistic shocks \citep{2009herb.book.....D}, allow us to consider
a model that is effectively slowly cooled.  Simulations typically have
more parameters than our current model
\citep{Peer:2004,Asano:2009,Daigne:2009}, and constraining those
models via spectral templates using data from {\it Fermi} may be
difficult.

\subsection{Blackbody Component}
\label{sec:model:bb}
The pure fireball scenario for GRB emission predicts that most of the
flux is from thermal emission
\citep{Goodman:1986,Paczynski:1986}. This is because as the jet
becomes optically thin at some photospheric radius, $r_{ph}$, it
releases radiation that has undergone many scatterings with the
optically thick electrons below the photosphere. We model this
emission as a blackbody
\begin{equation}
  \label{eq:blackbody}
  F_{BB}(\mathcal{E})\;=\;A \mathcal{E}^3 \frac{1}{ e^{\frac{\mathcal{E}}{kT}}-1}
\end{equation}
where A is the normalization and kT scales the energy of the
function. This is simplified thermal emission from the photosphere
that does not take into account the effects of relativistic broadening
that can produce a multi-color blackbody emerging from the photosphere
\citep{Beloborodov:2010,Ryde:2010,Peer:2011}. \citet{Ryde:2006} showed
that if it is assumed that the thermal component is emanating from the
photospheric radius of the jet, several properties about the blackbody
component are derivable. The cooling behavior is well predicted for a
thermal component. The temperature of the blackbody should decay as
$T\propto r_{ph}^{-2/3}$. If $\Gamma$ is assumed to remain constant
during the coasting phase of the jet then it can be shown that the
temperature should decay as $T \propto t^{-2/3}$ in time.  It has been
found observationally that the evolution of kT often follows a broken
power-law trend with the index below the break averaging to $\sim
-$2/3 \citep{Ryde:2009}. Finally, a true blackbody has a well defined
relation between energy flux and temperature:
\begin{equation}
   \label{eq:sbl}
  F_{BB}=N\sigma_{sb} T^4
\end{equation}
where $N$ is a normalization related to the transverse size
  of the emitting surface, $r_{ph}/\Gamma$
  \citep{Ryde:2005,Ryde:2009,Iyyani:2013}, and $\sigma_{sb}$ is the
  Stephan-Boltzmann constant.

The photospheric radius and the transverse size of the photospheric
emitting region are also of great importance to understanding the geometry
and energetics of GRBs. In \citet{Ryde:2009}, a parameter
$\mathcal{R}$ 
\begin{equation}
  \label{eq:R}
  \mathcal{R}(t)\;\equiv\;\left(\frac{F_{BB}(t)}{\sigma_{sb}T(t)^4} \right)^{1/2} \propto \dover{r_{ph}}{\Gamma}.
\end{equation}
where $F_{BB}(t)$ is the time-dependent energy flux of the blackbody,
is used to track the outflow dynamics of the burst. The connection
between $\mathcal{R}$ and N from eq. \ref{eq:sbl} is established by
noting that $N = \mathcal{R}^2$. Thereby, only if $\mathcal{R}(t)$ is
constant would we expect to recover the relation established in
eq. \ref{eq:sbl}.  Several BATSE GRBs were found to have a power-law
increase of $\mathcal{R}$ with time. However, the connection between
$\mathcal{R}$, T, and $F_{BB}$ was difficult to establish because the
error on the data points was large. Understanding these connections is
essential to unmasking the structure and temporal evolution of GRB
jets.

\section{Time Resolved Analysis}

\label{sec:observe}

\subsection{Summary of Technique}
The GRBs in our sample were selected based on two criteria: large peak
flux and single-peaked, non-overlapping temporal structure. The GRBs
were binned temporally in an objective way described in Section
\ref{sec:observe:tbin} and spectral fits were performed on each time
bin using four different photon models (Band, Band+blackbody,
synchrotron, synchrotron+blackbody). When fitting synchrotron we
compared the fits of slow-cooling and fast-cooling synchrotron. In
many cases the fits from fast-cooling synchrotron completely
failed. From these spectral fits a photon flux lightcurve was
generated for each component and fitted with a pulse model to
determine the decay phase of the pulse. We describe each step in the
following subsections.

\subsection{Sample Selection}
\begin{figure}

 \centering

  \subfigure[]{

    \label{fig:lc:a}
    \includegraphics[scale=.70, angle=0]{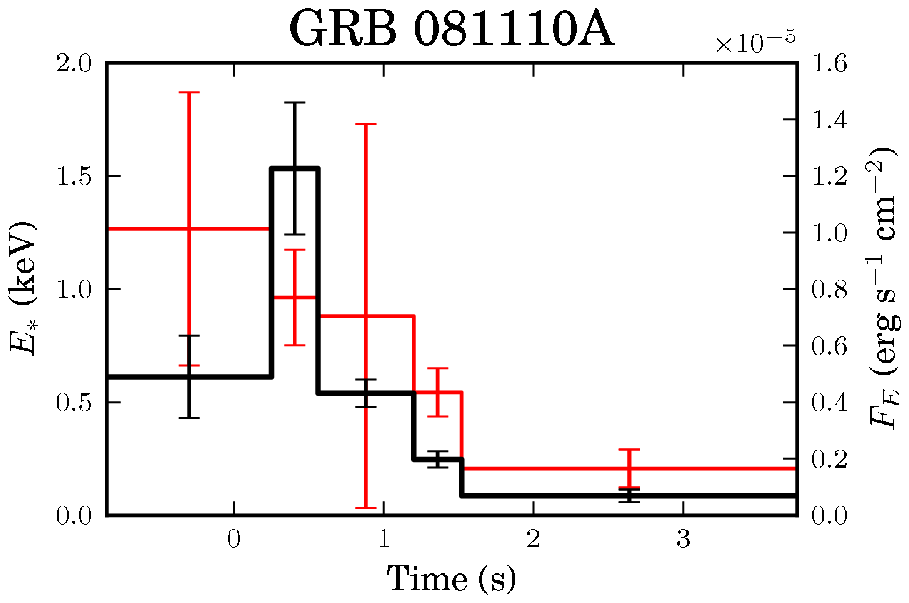}}\subfigure[]{
    \label{fig:lc:b}
    \includegraphics[scale=.70, angle=0]{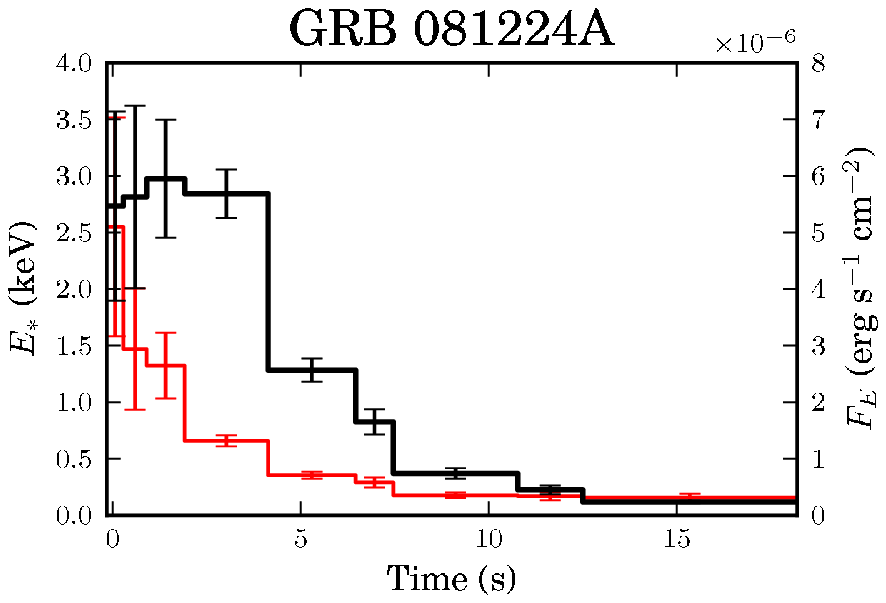}}
  \subfigure[]{
    \label{fig:lc:c}
    \includegraphics[scale=.70, angle=0]{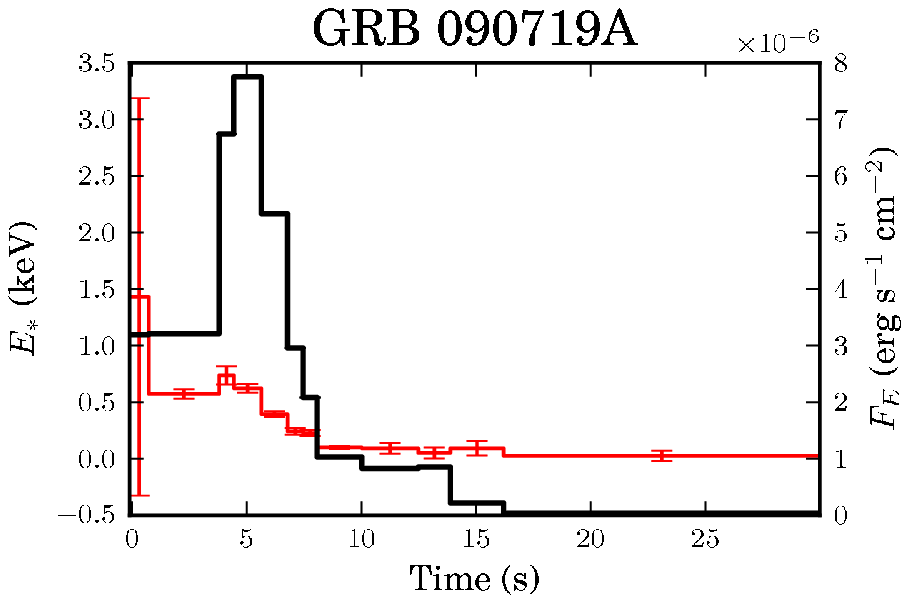}}\subfigure[]{
    \label{fig:lc:d}
    \includegraphics[scale=.70, angle=0]{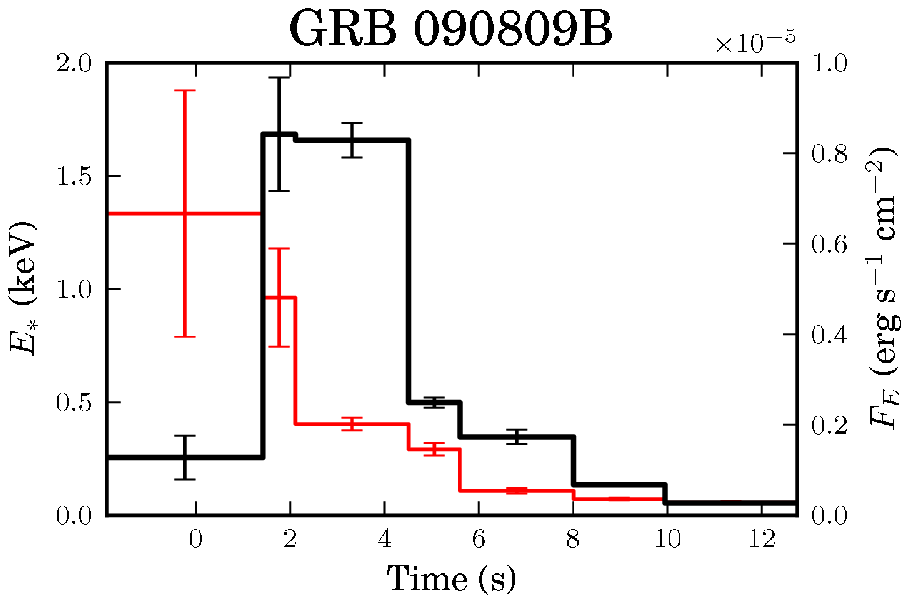}}
\subfigure[]{
    \label{fig:lc:e}
    \includegraphics[scale=.70, angle=0]{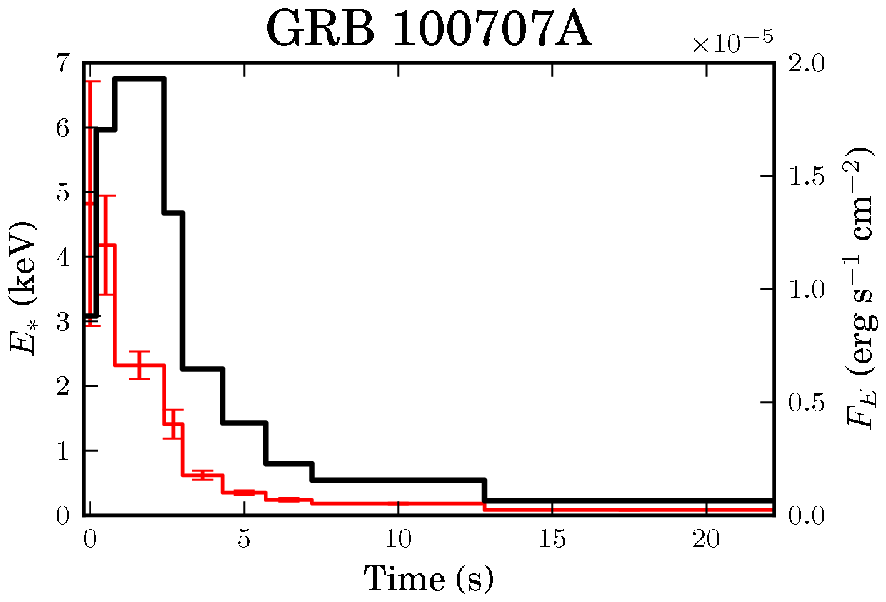}}\subfigure[]{
    \label{fig:lc:f}
    \includegraphics[scale=.70, angle=0]{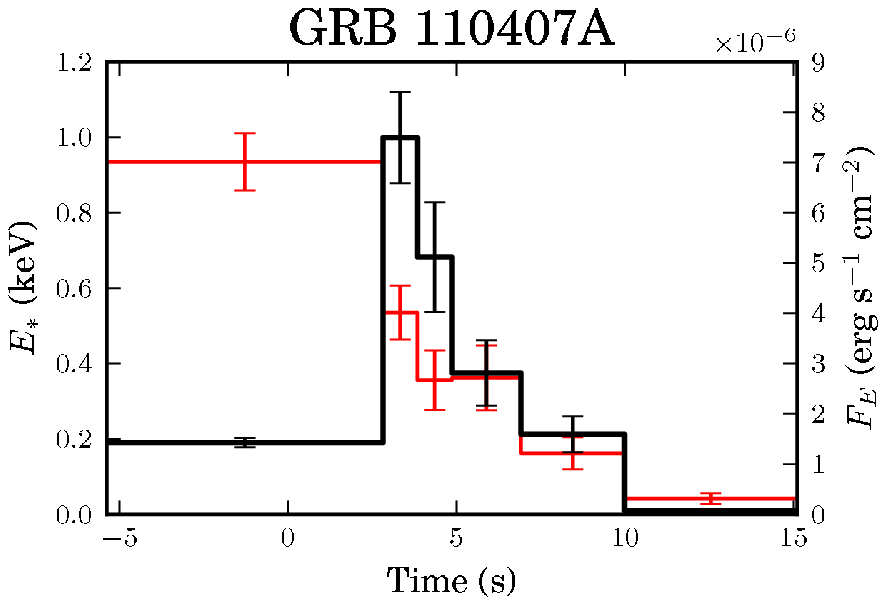}}
\subfigure[]{
    \label{fig:lc:g}
    \includegraphics[scale=.70, angle=0]{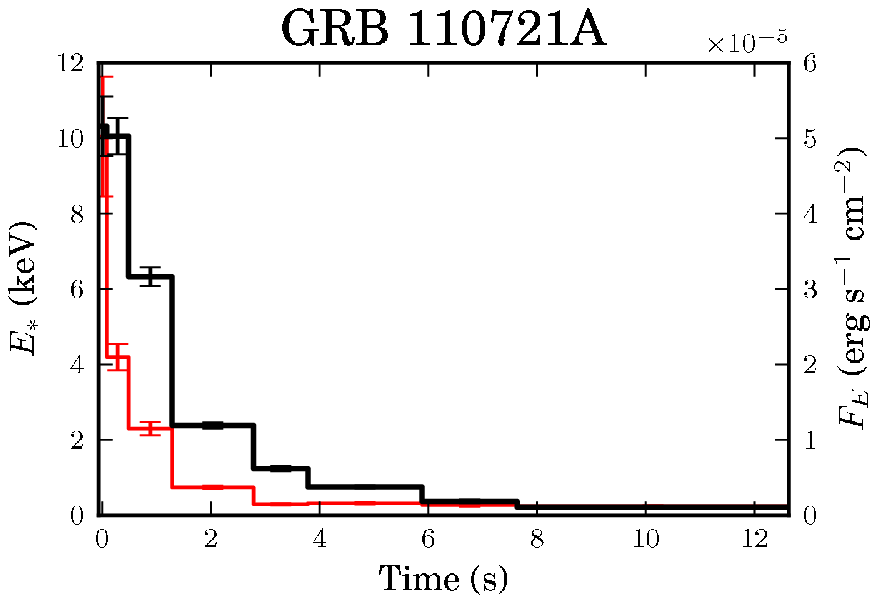}}\subfigure[]{
    \label{fig:lc:h}
    \includegraphics[scale=.70, angle=0]{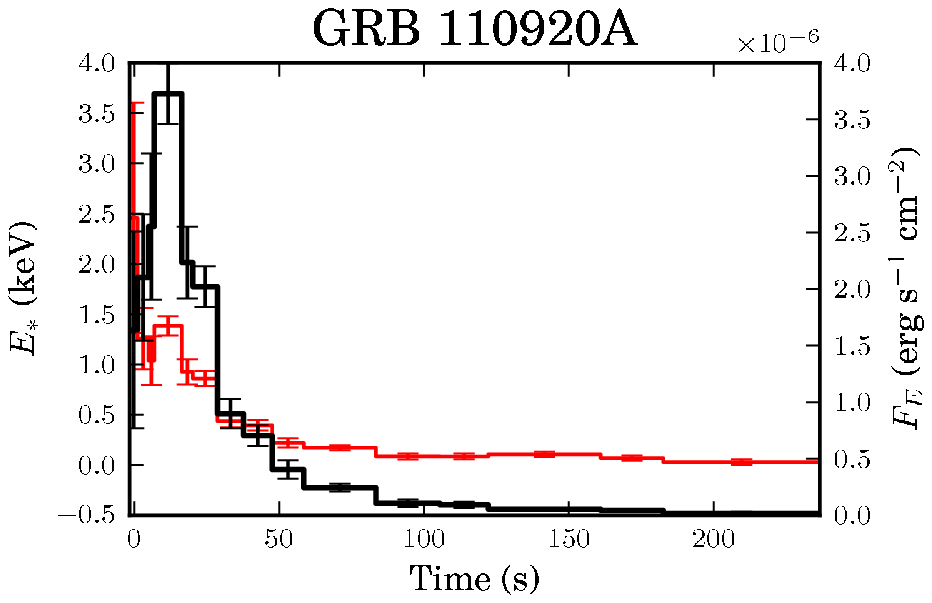}}

  \caption{The energy flux lightcurves of the synchrotron component
    for the entire sample (\emph{black} curve). The integration range
    is from 10 keV - 40 MeV for all GRBs except GRB 081224A and GRB
    110721A which are from 10 keV - 300 MeV. Superimposed is the
    slow-cooled synchrotron $E_p$ (see eq. \ref{eq:ergcsynch})
    (\emph{red} curve) demonstrating the hard to soft evolution of the
    bursts. }

  \label{fig:lc}

\end{figure}

To fully constrain the parameters of the fitted models, we selected
GRBs with a requirement that the peak flux be greater than 5 photons
s$^{-1}$ cm$^{-2}$ between 10 keV and 40 MeV. It is important for our
GRBs to have a simple, single-peaked lightcurve structure to avoid the
overlapping of different emission episodes. This facilitates
  the identification of distinct evolutionary trends in the physical
  parameters for the emission region. While we cannot be sure that a
weaker emission episode does not lie beneath the main peak, the bursts
we selected have no significant additional peak during the rise or
decay phase of the pulse. These two cuts left us a sample of eight
GRBs: GRB 081110A \citep{GRB081110A}, GRB 081224A \citep{GRB081224A},
GRB 090719A \citep{GRB090719A}, GRB 090809B, GRB 100707A \citep{
  GRB100707A}, GRB 110407A, GRB 110721A, \citep{GRB110721A} and, GRB
110920A (Figure \ref{fig:lc}). GRB 081224A and GRB 110721A were both
analyzed including the new LAT Low-Energy (LLE) data that provides a
high effective area above 30 MeV for the analysis of short-lived
phenomena, thanks to a loosened set of cuts with respect to LAT
standard classes \citep{Vero:2010, Ack:2012a}. This data selection
bypasses the typical photon classification \citep{Ackerman:2012} tree
and includes events that would normally be excluded but can be
selected temporally when the signal to background rate is high, such
as with GRBs. GRB 081224A had very little data above 30 MeV but the
LLE data helped to constrain the spectral fits. From this sample, five
GRBs (GRB 081224A, GRB 090719A, GRB 100707A, GRB 110721A, and GRB
110920A) had blackbody components that were bright enough to analyze
(Table \ref{tab:grbs}).

\subsection{Time Binning}
\label{sec:observe:tbin}
In order to bin the count data, the time bins for spectral analysis of
each GRB were chosen using Bayesian blocks
\citep{Scargle:1998}. Bayesian blocks define the time bins by looking
for significant changes in the data count rate that define change
points based on a Bayesian prior. For GBM data, we combined the TTE
data from the brightest NaI detector with the TTE data from the BGO
detector and ran a Bayesian block algorithm to find the times of the
change points. We found that a prior of 8 gave a good balance between
time resolution and having enough counts to perform spectral
analysis. These change points were mapped to the other detectors used
in the analysis. This method only accounts for changes in counts and
therefore could inadvertently combine bins with different spectral
shapes.

\subsection{Spectral Analysis}
For spectral fits we used the RMFIT
ver4.1\footnote{http://fermi.gsfc.nasa.gov/ssc/data/analysis/user/}
software package developed by the GBM team. Fitting the synchrotron
photon model requires a custom module developed and used in B11. Each
time bin was fit with one of the four spectral models mentioned
above. We fit the physical models to compare the validity of each one
against the other and the Band function to try and understand how the
Band function parameters correlate with the best fit physical
model. If the addition of blackbody component did not make a
significant improvement of at least 10 units of C-stat
\citep{Arnaud:2011} for any time bins of a particular GRB, then we did
not include the blackbody component in the analyzed fits for that
burst. C-stat is a Poisson likelihood with constant offset. However,
near the end of the prompt emission in some GRBs, the blackbody
component becomes weak but has spectral evolution consistent with more
significant time bins in the burst. The spectral parameters of the
blackbody in those bins were included even though they contributed
large error bars to some quantities. We checked with simulations that
this cut was sufficient to identify a significant addition of a
blackbody to the fit model.

\section{Results}
\label{sec:results}
\subsection{Test of Slow-Cooling Synchrotron}
\label{sec:results:scs}

\begin{figure}

  \centering

 \includegraphics[scale=1, angle=0]{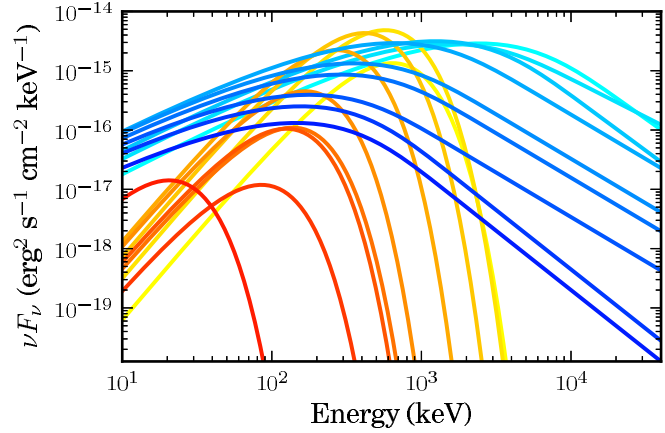}

 \caption{The spectral evolution of GRB081224A is an example of the
   typical evolution observed for the entire sample.  The synchrotron
   (from \emph{light blue} to \emph{dark blue}) and blackbody (from
   \emph{yellow} to \emph{red}) both evolve from hard to soft peak
   energies with time. For this GRB, the high-energy power-law
   corresponding to the electron spectral index does not evolve
   significantly over the duration of the burst.}

  \label{fig:specEvo}

\end{figure}

\begin{figure}

 \centering

 \subfigure[]{
   \label{fig:counts:a}
  \includegraphics[scale=.25, angle=0]{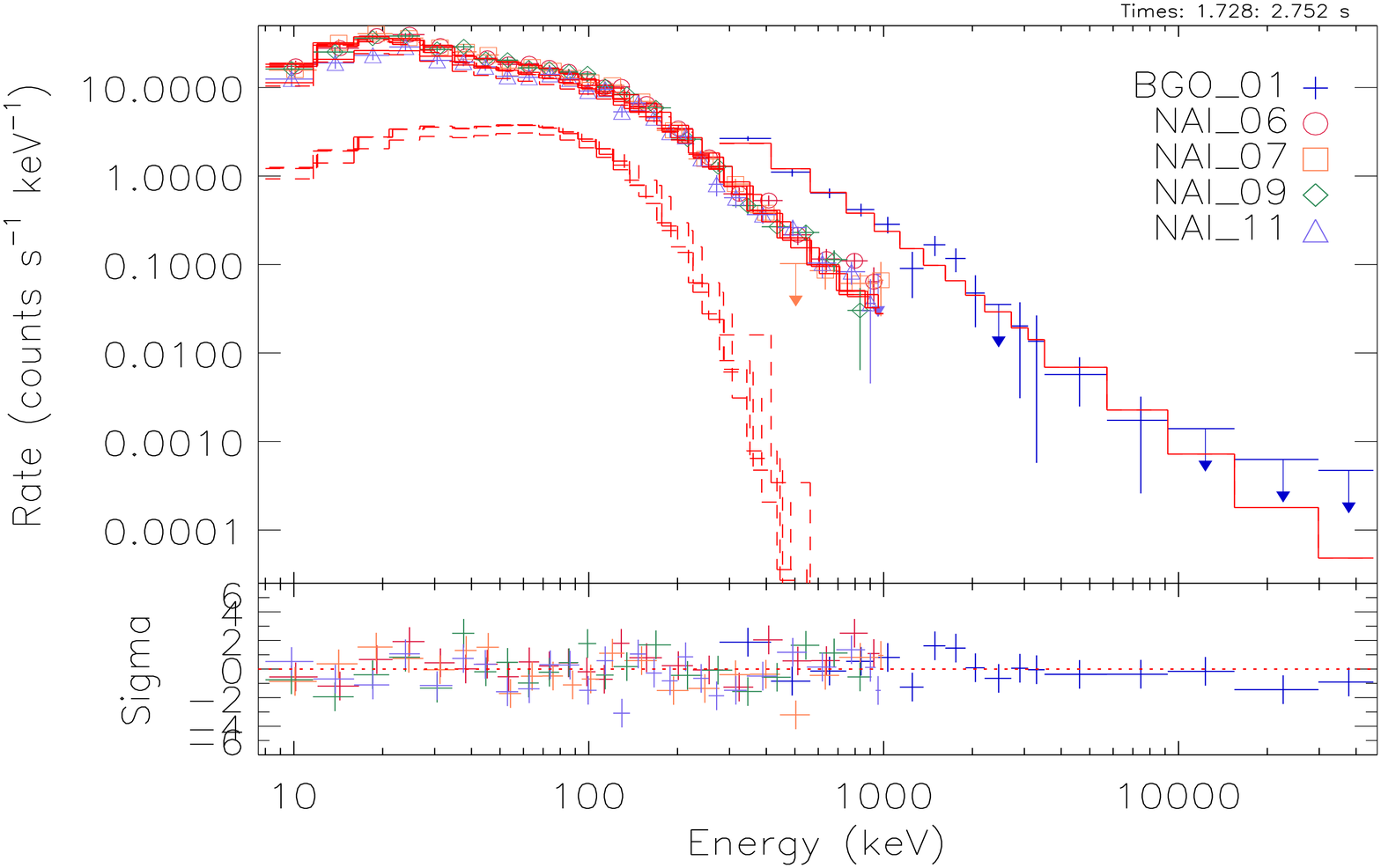}
}\subfigure[]{
   \label{fig:counts:b}
   \includegraphics[scale=.44, angle=0]{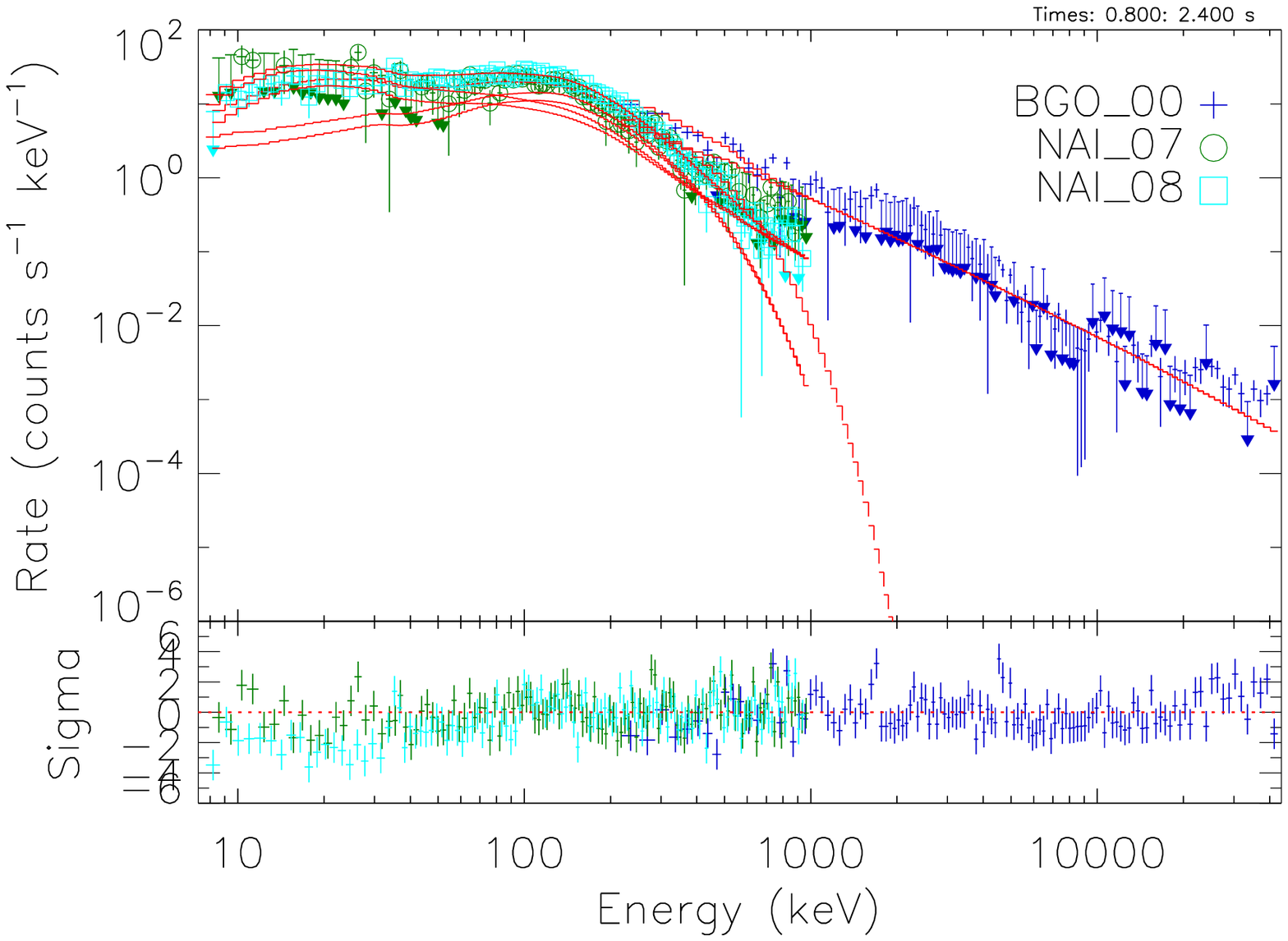}
}

\caption{A time bin of GRB 110721A (left panel) and GRB 100707A (right
  panel) demonstrating typical count spectra from the sample. Two
  extreme cases are shown: a subdominant and dominant blackbody
  component. The response has been convolved with synchrotron (eq.
  \ref{eq:synch_flux}) and a blackbody to produce counts. The
  residuals from the fits indicate that the model is fitting the data
  well.}

\label{fig:counts}

\end{figure}

In nearly all cases the synchrotron or synchrotron+blackbody model
produced a fit with a comparable or better C-stat than the Band
function. The GRBs exhibited hard to soft spectral evolution (Figure
\ref{fig:specEvo}) for both components. From these fits we can derive
several interesting properties of the bursts. The results of fitting
the non-thermal part of each time bin in our sample with slow-cooled
synchrotron indicate that this model can indeed fit the data well. The
C-stat fit statistic per degree of freedom was at or near 1 for most
time bins. The spectral fit residuals cluster around zero, with no
deviations at low-energy that might indicate the presence of an
additional power-law component (Figure \ref{fig:counts}). The
residuals are below 4$\sigma$ for the entire energy range. As an
example, the fit C-stats for GRB 100707A and GRB 110721A are shown in
Tables \ref{tab:grb1c} and \ref{tab:grb2c} respectively. They show
that the slow-cooled synchrotron model fits the data as well as the
Band function when a blackbody is included in both cases. The fit
C-stats for fast-cooled synchrotron are shown for GRB 110721A to
compare all three non-thermal models. These results imply that
slow-cooled synchrotron is a viable model for GRB prompt emission. We
cannot claim it provides a better fit to the data than other untested
models and we will investigate and compare other physical models in
future work

\begin{figure}

 \centering

  \includegraphics[scale=1, angle=0]{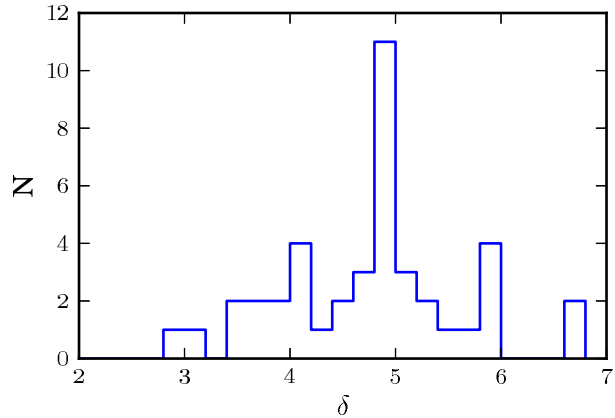}

  \caption{The distribution of electron indices from the slow-cooling
    synchrotron fits. Only indices that were constrained are
    plotted. The distribution is broad but centered at $\delta=$5
    which is much steeper than expected from simple relativistic shock
    acceleration.}

  \label{fig:index}

\end{figure}

An important parameter constrained in these fits is the electron
index, $\delta$, of the accelerated power-law. The canonical value for
diffusive acceleration at ultra-relativistic, parallel shocks is
$\delta$=2.2 \citep{Kirk:1987,Kirk:2000}. The distribution of
constrained $\delta$'s (Figure \ref{fig:index}) is broad and centered
around $\delta=5$ (i.e., $\beta = 3$).  This steep index
could provide clues for the structure and magnetic turbulence spectrum
of the shocks. \citet{Baring:2006}, \citet{Ellison:2004} and
\citet{Baring:2012} show that shock speed, obliquity, and turbulence
all have a strong effect on the electron spectral index of the
accelerated electrons. Steeper indices correspond to increasing shock
obliquity in superluminal shocks. Fit models which are
built from the electron distribution such as the one used in this work
enable a direct diagnostic of the GRB shock structure.

\subsection{Test of Fast-Cooling Synchrotron}
\label{sec:fast}
\begin{figure}

 \centering

  \includegraphics[scale=.7, angle=0]{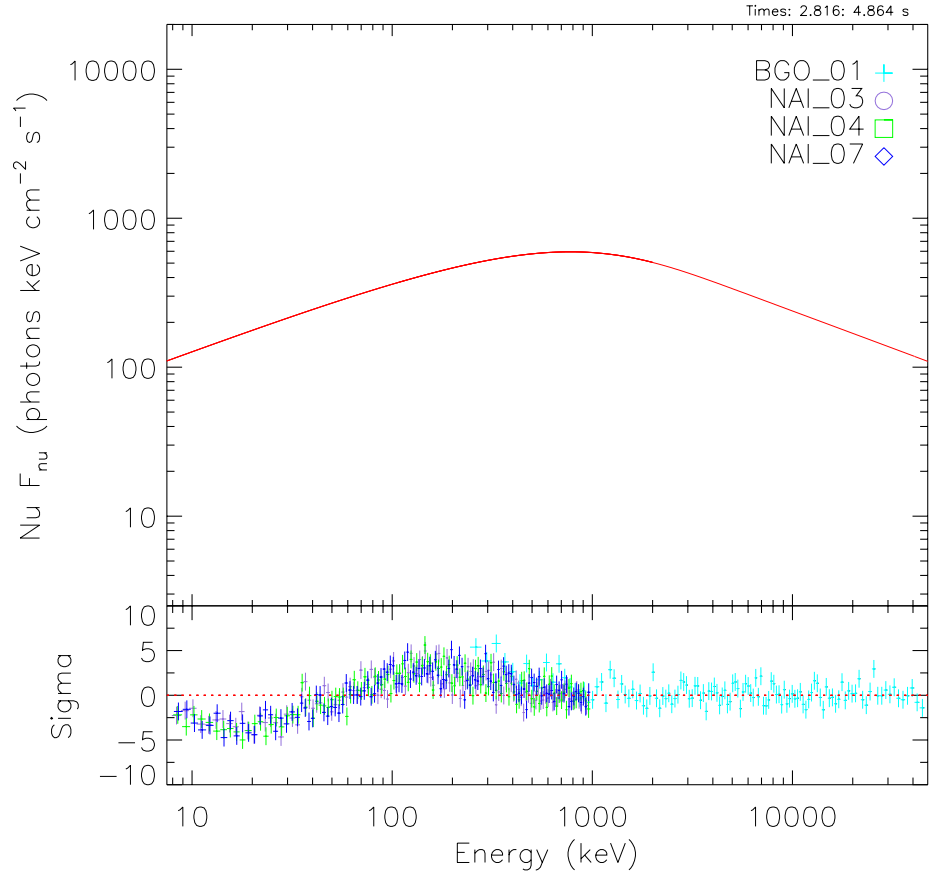}

  \caption{The fast-cooled synchrotron fits are poor for nearly all of our
    sample because none of the spectra have a low-energy
    index as steep as $-$3/2. Therefore the fast-cooled synchrotron
    spectrum is too broad around the $\vFv$ peak as shown in
    this example spectrum.}

  \label{fig:fastS}

\end{figure}

\begin{figure}

 \centering

  \includegraphics[scale=1, angle=0]{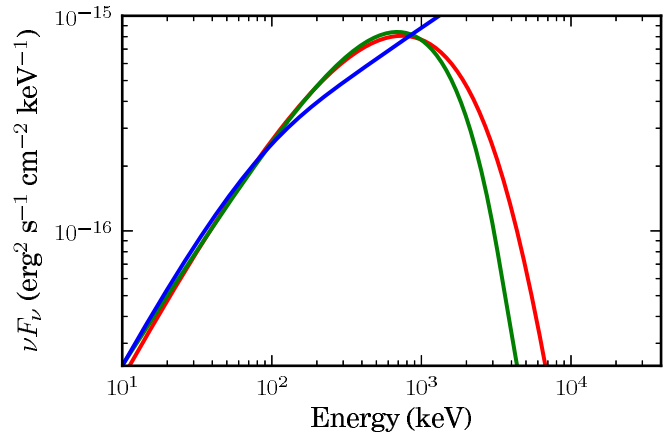}

  \caption{An example time bin of GRB 110407A comparing the fitted
    $\vFv$ spectra of the Band Function (\emph{green}),
    slow-cooled synchrotron (\emph{red}), and fast-cooled synchrotron
    (\emph{blue}). While the Band function and slow-cooled synchrotron
    fits resemble each other, the fast-cooled synchrotron fit is only
    able to fit the low-energy part of the spectrum. Because
    fast-cooled synchrotron has an index of $-$2/3 below the cooling
    frequency, the fitting engine pushes the value of E$_*$ very high
    to fit the low-energy part of the spectrum resulting in a $-$3/2
    index near the $\vFv$ peak. The high-energy power-law of
    the fast-cooling synchrotron spectrum is pushed out of the data
    energy window.}

  \label{fig:fastComp}

\end{figure}

In order to see if any spectra were consistent with the fast-cooling
synchrotron spectrum, we implemented a fast-cooled synchrotron model
where the electrons were distributed according to the broken power-law
in eq. \ref{eq:necool}. These apply to the non-thermal particles
initially accelerated in the shock neighborhood, that convect and
diffuse out of this zone and subsequently cool radiatively in a much
larger volume. As with the slow-cooling fits, $\gammaMin$ was held
fixed to 900. Several spectra were tested and all resulted in very
poor fits regardless of whether the low-energy index found with the
Band function was much harder than $-$3/2 (Figure~\ref{fig:fastS} and
Table \ref{tab:fast}). This is due to the broad spectral curvature of
the fast-cooled spectrum around the $\nu F_{\nu}$ peak. The broken
power-law nature of the electron distribution is smeared out by the
synchrotron kernel and cannot fit the typical curvature of the GBM
data. In fact, the fast-cooling synchrotron spectrum has a spectral
index of $-$2/3 below the $\gamma_c$ which we have fixed at 1. The
fitting algorithm increased $E_*$ to high values to align the $-$2/3
index with the data, which resulted in poor fits (Figure
\ref{fig:fastComp}). Even when a blackbody is present in the bursts,
fast-cooled synchrotron is not a good fit to the non-thermal part of
the spectrum (Table \ref{tab:grb2c}). The lack of GRBs with low-energy
indices as steep as $-$3/2, additionally disfavors fast-cooled
synchrotron as the non-thermal emission component in GRB spectra.

\subsection{Synchrotron vs. Band}
\label{sec:results:bvs}
\begin{figure}

  \centering

  \includegraphics[scale=1, angle=0]{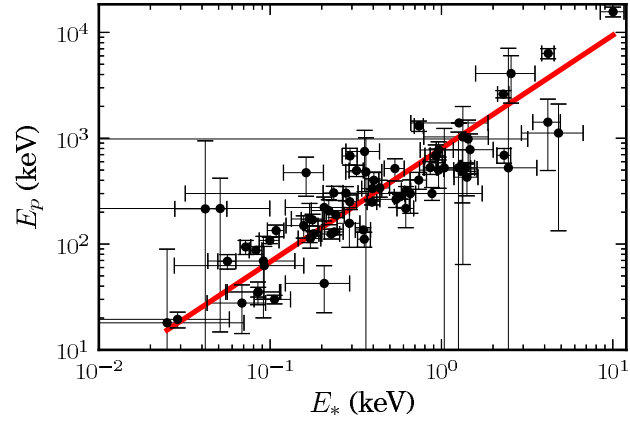}

  \caption{Derived values of the parameter E$_{\rm p}$ (obtained using
    the Band function to fit GRB spectra), versus $E_*$ (obtained using
    an optically-thin non-thermal synchrotron to fit GRB spectra).}

  \label{fig:EpEc}

\end{figure}

The Band function has been used in the literature as a proxy for
distinguishing among non-thermal emission mechanisms. The predicted
non-thermal emission of GRBs is typically characterized as a
smoothly-broken power-law with the high-energy spectral index related
to the index of accelerated electrons and the low-energy index related
to the radiative emission process. Therefore, fitting a Band function
to the emission spectrum of a GRB {\em should} serve as a diagnostic
of the radiative process responsible for the
emission. \citet{Preece:1998} examined the BATSE GRB catalog and
looked at the distribution low-energy indices from Band function
fits. They found that the distribution peaked at $\alpha\approx -1$
and that 1/3 of the fitted spectra had low-energy indices too hard for
synchrotron radiation. The assumption is that the Band function's
shape approximates synchrotron but has an added degree of freedom in
the low-energy index. However, the Band function has a broader range
of curvatures around the $\vFv$ peak allowing it the possibility to
deviate from the shape of synchrotron above and around the $\vFv$
peak. The synchrotron $\vFv$ peak is $\propto\gammaMin^2 B\propto
E_*$, leading to the relation between Band and synchrotron models
E$_{\rm p}\propto E_*$. This relationship is easily recovered from our
sample (Figure \ref{fig:EpEc}). Direct comparison of the quality of
the fits using Band and the synchrotron model is not the goal of this
study. Both Band and the synchrotron model fit the data well with
their respective fit residuals not deviating more than 4$\sigma$ and
centered around zero (Figure \ref{fig:counts}). It is important to
stress that the questions being asked are does the synchrotron model
fit the data?  and, what temporal evolution do the synchrotron
parameters undergo?

For all GRBs in our sample that include both a blackbody and
non-thermal component we compare the photon flux (photons s$^{-1}$
cm$^{-2}$) lightcurves (integrated from 10 keV - 40 MeV) derived from
synchrotron fits with those derived from Band fits (Figure
\ref{fig:fluxComp}). It is seen that while both methods recover the
same total flux, the flux from the individual components is much
better constrained when using the synchrotron model for the
non-thermal component. This is due to the pliability of the Band
function below E$_{\rm p}$ that is not afforded to the synchrotron
model.

The C-stat fit values for the synchrotron model loosely correlate with
the value of Band $\alpha$ found by fitting the same interval with the
Band function. When Band $\alpha$ was much harder than zero, the
synchrotron fit was poor and typically required adding blackbody to
fit the data. The flexibility of the Band function with its low-energy
power law creates the possibility that the index alpha of that power
law will not accurately measure the true slope if E$_{\rm p}$ is too
close to the low-energy boundary of GBM data. Simulated spectra using
the Band function were created with a grid in both Band $\alpha$ and
E$_{\rm p}$ to ensure that low values of E$_{\rm p}$ do not affect the
reconstruction of Band $\alpha$ in our fits. It was found that Band
$\alpha$ could be accurately measured when E$_{\rm p}$ was as low as
$\sim$20 keV. While the asymptotic value of synchrotron is $-$2/3,
fitting the photon model with an empirical function like Band with a
slightly different curvature could result in measured low-energy
indices that are different. To measure this effect, simulated
synchrotron spectra with different $E_*$ were fit with the Band
function. The Band $\alpha$ showed a slight dependence on the
synchrotron peak; moving to softer values for lower $E_*$. The
distribution of fitted Band $\alpha$ values from these simulations
centered around $-$0.81 $\pm$ 0.1, a slightly softer value than $-$2/3
which may explain the clustering of Band $\alpha$ at $-$0.82 in the
GBM spectral catalog \citep{Gold:2012a} if a majority of the
non-thermal spectra are the result of synchrotron emission.

\subsection{High-Energy Correlations}
\label{sec:results:hec}
There is a well-known spectral evolution in GRB pulses of $E_{\rm
  peak}$ evolving from hard to soft (see Figures \ref{fig:lc} and
\ref{fig:specEvo}). This leads to two time-resolved correlations
between hardness (measured as E$_{\rm p}$) and flux
\citep{Golenetskii:1983,Liang:1996,Ghisellini:2010}. \citet{Liang:1996}
(hereafter LK96) showed that the hardness intensity correlation (HIC)
which relates the instantaneous energy flux $F_{E}$ to spectral
hardness and can be defined as
\begin{equation}
  F_E\;=\;F_0\left(\frac{E_{\rm p}}{E_{{\rm p},0}} \right)^q \;,
\end{equation}
where $F_0$ and E$_{\rm p,0}$ are the initial values at the start of the
pulse decay phase and $q$ is the HIC index. \citet{Ryde:2001} found
that 57\% of a sample of 82 BATSE GRBs were consistent with this
relation. The second relation is the hardness-fluence correlation
(HFC) which relates hardness to the time-running fluence of the
GRB. Time-running fluence, $\Phi(t)$, is defined as the cumulative,
time-integrated flux of each time bin in a GRB. The HFC is expressed
as
\begin{equation}
  \label{eq:hfc}
  E_{\rm p}\;=\;E_{\rm p,0}e^{-\Phi(t)/\Phi_0}\;,
\end{equation}
where $\Phi_0$ is the decay constant. LK96 noted that this equation is
similar to the form of a confined radiating plasma. This should not be
the case for optically-thin synchrotron. Upon differentiating
eq. \ref{eq:hfc} it becomes apparent that the change in hardness is
nearly equal to the energy density:
\begin{equation}
  \label{eq:plasma}
  -\frac{dE_{\rm p}}{dt}\;=\;-\frac{F_{\nu}E_{\rm p}}{\Phi_0}\;\approx\;-\frac{F_E}{\Phi_0}.
\end{equation}
The HFC could be the result of a confined plasma with a fixed number
of particles cooling via $\gamma$-radiation as proposed by LK96.
Since these relations are only applicable during the decay
phase of a pulse the value of $T_{max}$ (the time of the peak flux)
from the pulse fit of each GRB is used as the initial point for $F_0$
and $E_{\rm p,0}$.

The use of a hardness indicator is somewhat ambiguous. Historically,
the ratio of counts in low and high-energy channels was used as a
hardness measure. This has an advantage of being model-independent but
suffers from the lack of information associated with the instrument
response. High-energy photons can scatter in the detector and not
deposit their full energy thereby artificially lowering the hardness
ratio. LK96 used the Band function E$_{\rm p}$ to compute
hardness, which as a deconvolved quantity is less instrument-dependent
but introduces a model dependence. We take this approach for both the
Band and synchrotron model fluxes. For synchrotron we use the $E_*$
parameter as our hardness indicator. This is justified by the
relationship between $E_*$ and E$_{\rm p}$ (see
Sec. \ref{sec:results:bvs} and Figure \ref{fig:EpEc}).

\begin{figure}

  \begin{center}
    \subfigure[]{
      \label{fig:Epcor:a}
      \includegraphics[scale=1, angle=0]{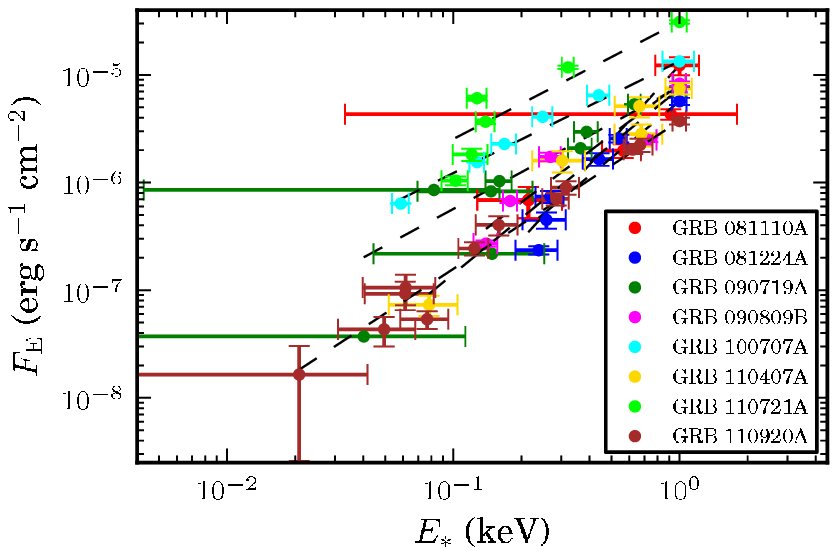}
    }    \subfigure[]{
      \label{fig:Epcor:b}
      \includegraphics[scale=1, angle=0]{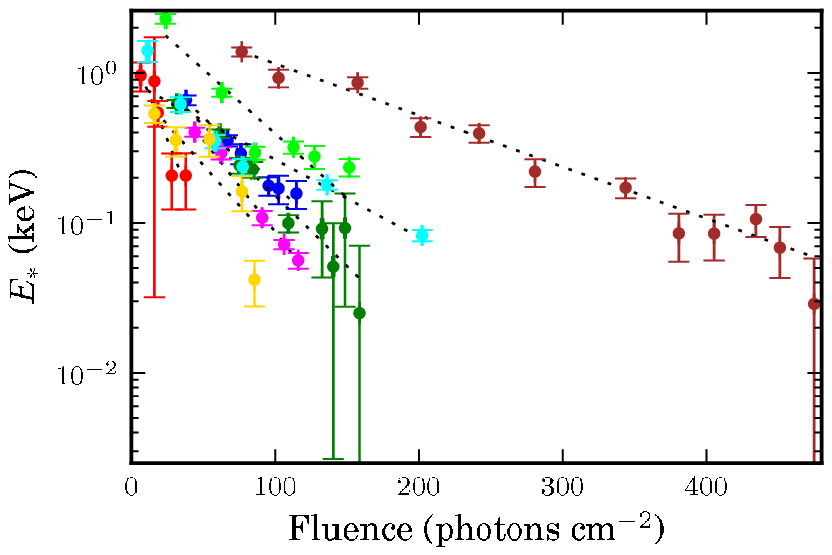}
    }
  \end{center}  

  \caption{The non-thermal emission of all of the bursts in the sample
    loosely follow $F_E$-$E_p$ and $E_p$-fluence relations. See Table
    \ref{tab:cor} for the numerical results. }

  \label{fig:Epcor}

\end{figure}

We compute the HIC and HFC for the synchrotron fits for each GRB in
our sample (Figure \ref{fig:Epcor} and Table \ref{tab:cor}). All the
GRBs seemed to follow the HIC to some extent. We find that the HIC
index for $E_*$ ranges between $\approx$1-2. When using the Band
function E$_{\rm p}$ as a hardness indicator, it has been suggested
that E$_{\rm p} \propto {\rm L}^{1/2} \propto F_{E}^{1/2}$, if one
assumes that the emission is due to synchrotron and that only $\Gamma$
changes while the other internal properties remain (however unlikely)
the same \citep{Ghisellini:2010}. Decay behavior due to light
travel-time effects of a briefly illuminated relativistic spherical
shell varies according to $E_{p}\propto L^{1/3}$, that is, $q = 3$
\citep{Kumar:2000,Dermer:2004,Genet:2009}, whereas GRB observations
here show $L\propto F \propto E_{p}^{1.1}$ -- $E_{p}^{2.3}$ (Table
\ref{tab:cor}).  Evolution of internal parameters that would explain
the observed correlations is an open question. The synchrotron fits
seem to obey the HFC fairly well. Owing to the large errors in the
Band flux, fits with synchrotron are more consistent with the HFC and
HIC than those with Band. The deviations of the data from the expected
synchrotron HIC may be due to the fact that there are overlapping
pulses under the main emission that alter the decay profile. In
addition, the use of Bayesian blocks to select time bins ignores
spectral evolution. If bins with very different E$_{\rm p}$ are
combined then it could affect the the HIC and HFC data.

\subsection{Blackbody component}
\label{sec:results:bb}
\begin{figure}

  \centering

   \subfigure[]{
    \label{fig:fluxComp:a}
    \includegraphics[scale=.8, angle=0]{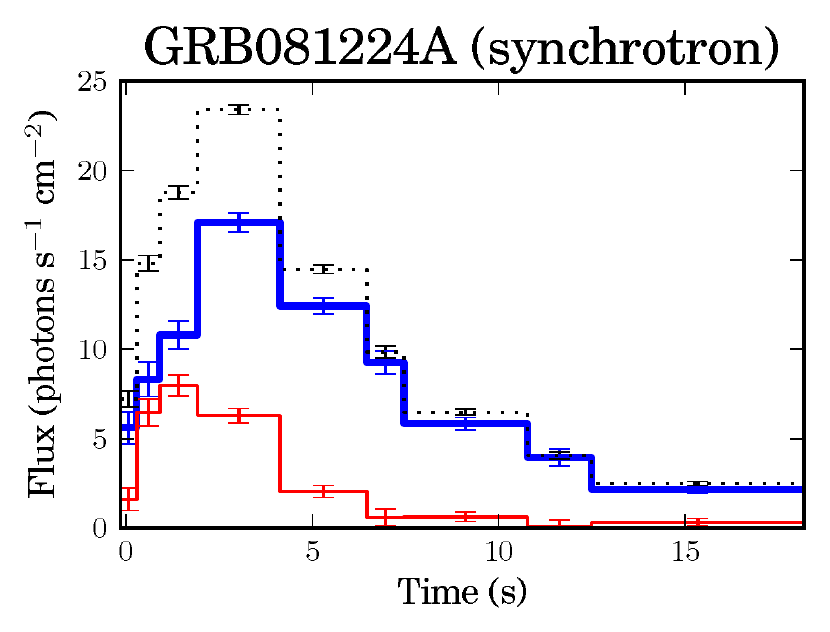}}\subfigure[]{
    \label{fig:fluxComp:b}
    \includegraphics[scale=.8, angle=0]{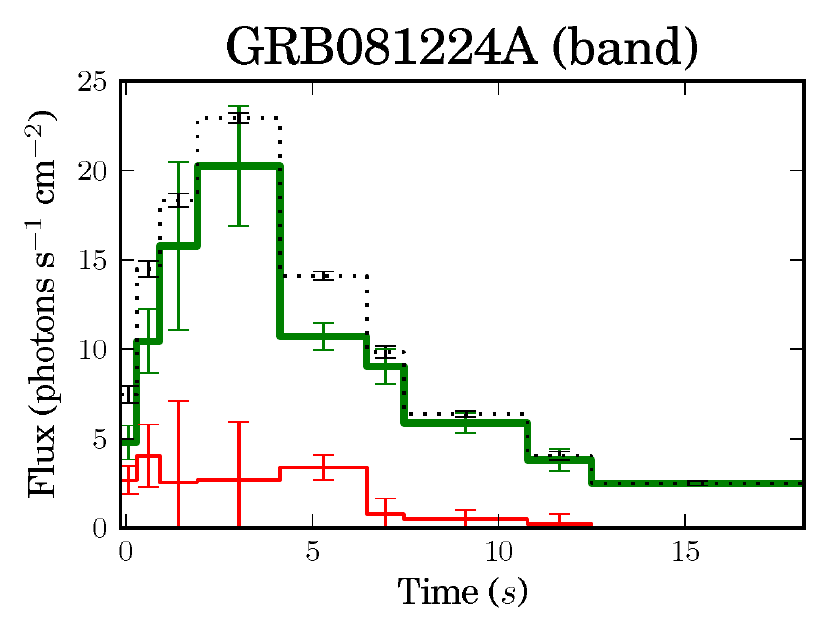}}
  \subfigure[]{
    \label{fig:fluxComp:c}
    \includegraphics[scale=.8, angle=0]{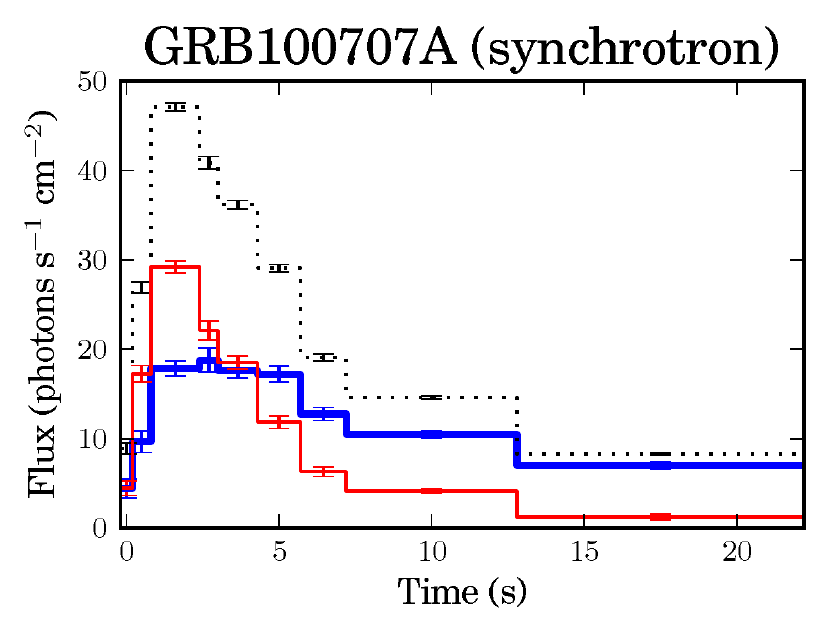}}\subfigure[]{
    \label{fig:fluxComp:d}
    \includegraphics[scale=.8, angle=0]{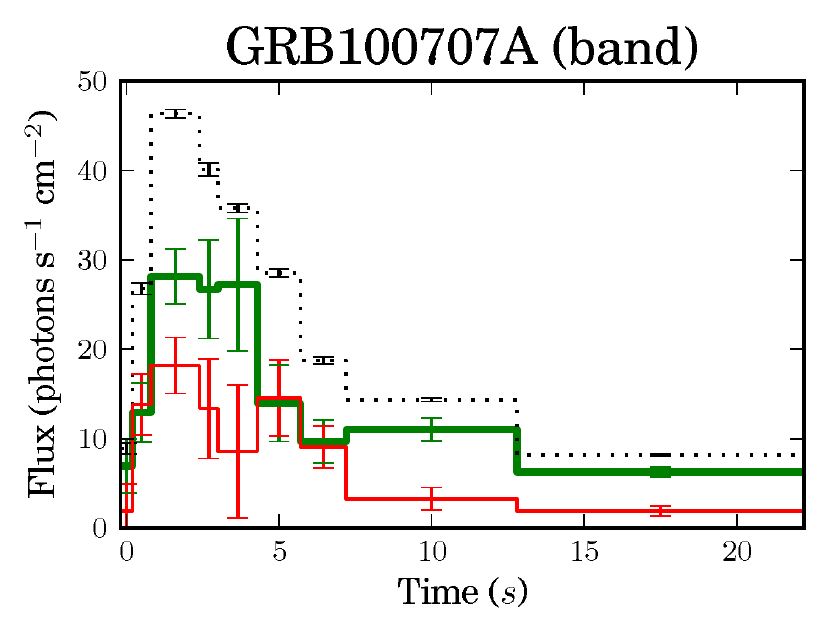}}
\subfigure[]{
    \label{fig:fluxComp:e}
    \includegraphics[scale=.8, angle=0]{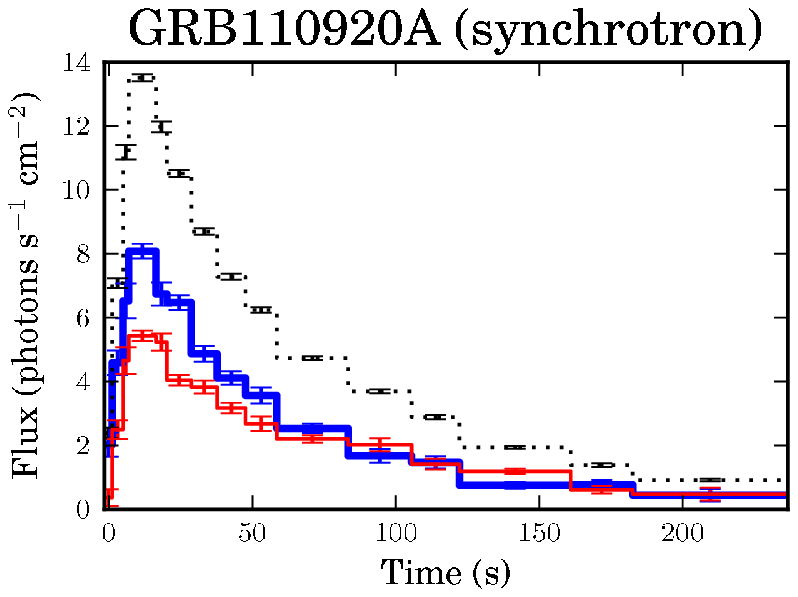}}\subfigure[]{
    \label{fig:fluxComp:f}
    \includegraphics[scale=.8, angle=0]{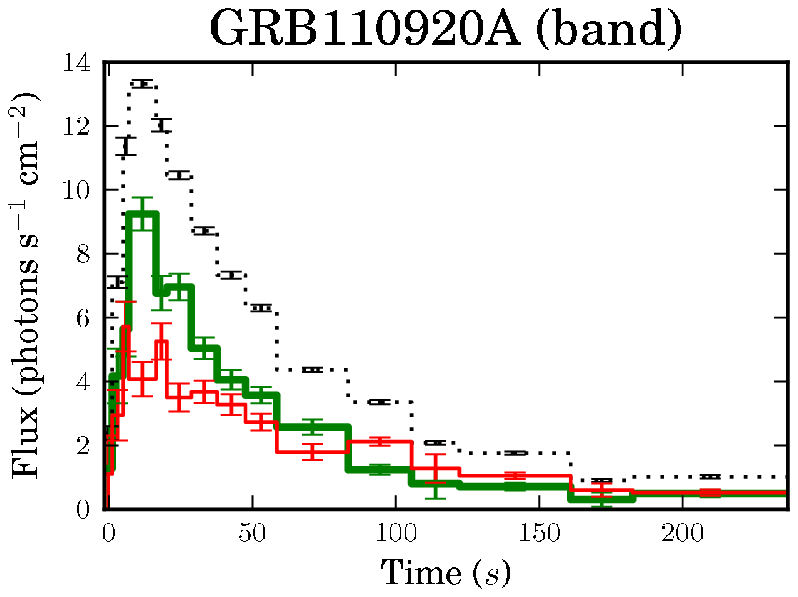}}

  \caption{A subset of flux lightcurves illustrating both the temporal
    structure of the different components and the advantages of using a
    physical model to deconvolve the detector response. The left
    column contains the lightcurves using synchrotron (\emph{blue
      thick line}) and blackbody (\emph{red thin line}) while the
    right column contains the lightcurves made from using the Band
    function (\emph{green thick line}) and blackbody (\emph{red thin
      line}). The total flux lightcurve (\emph{black dotted line}) of
    both approaches are the same. The components have a very simple
    and constrained evolution when using synchrotron as the
    non-thermal component. This is potentially indicative that
    synchrotron is the actual emission mechanism and the response is
    being properly deconvolved. In contrast, the lightcurves where the
    Band function is used have large errors and the blackbody does not
    have a consistent evolution. (See the online version for color)}

  \label{fig:fluxComp}

\end{figure}

For most of the spectra in our sample, the blackbody's $\vFv$ peak is
below the $\vFv$ peak of the non-thermal component. There is sometimes
a much larger change in C-stat between fits with synchrotron and
synchrotron+blackbody than those of Band and Band+blackbody owing to
the fact that the Band function has more freedom in the shape below
E$_{\rm p}$ (Tables \ref{tab:grb1dc} and
\ref{tab:grb2dc}). Simulations of both Band and slow-cooling
synchrotron were used to find the significance of adding a blackbody
to the spectrum. It was found that even when the difference in C-stat
between Band and Band+blackbody was greater than the difference
between synchrotron and synchrotron+blackbody, the statistical
significance in the goodness of fit after the addition of the
blackbody is high if not greater for the synchrotron+blackbody model
for many cases. Computational time limits kept us from checking if the
significance reached 5$\sigma$.  We now focus on the blackbody
component that is found in the synchrotron+blackbody fits. The
blackbody appears to have a separate temporal structure from the
non-thermal component, typically peaking earlier in time and decaying
before the non-thermal emission. (Figure \ref{fig:fluxComp}).

The form of the blackbody used in this work (eq. \ref{eq:blackbody})
is simplified and therefore will likely only approximate the true form
of thermal emission from a GRB photosphere. Since the blackbody is
weaker than the non-thermal (synchrotron) component in the spectrum,
the effects of a broadened and more realistic relativistic blackbody
are masked and would only slightly affect the fit when
combined with the synchrotron model. However, in the case of GRB
100707A, the blackbody is very bright (Figure \ref{fig:fluxComp:c})
and subtle changes in actual shape of the photospheric emission become
more apparent. This is reflected in the C-stat values in Table
\ref{tab:grb1c}. The Band function combined with the standard
blackbody is a better fit than when using synchrotron as the
non-thermal component. In this case, the Band function $\alpha$ is
still very hard indicating that the Band function is making up for
additional flux that the blackbody is not taking into account. To test
this hypothesis, we fit the synchrotron model along with an
exponentially cutoff power-law to mimic a modified blackbody. We found
that the fits were as good as the Band function combined with
blackbody fits indicating that when the blackbody is bright compared
to the non-thermal emission a more detailed model of the photospheric
emission is needed to fit the thermal part of the spectrum.

\begin{figure}

  \begin{center}

    \subfigure[]{
      \label{fig:kTcor:a}
      \includegraphics[width=8.5cm,height=6.5cm, angle=0]{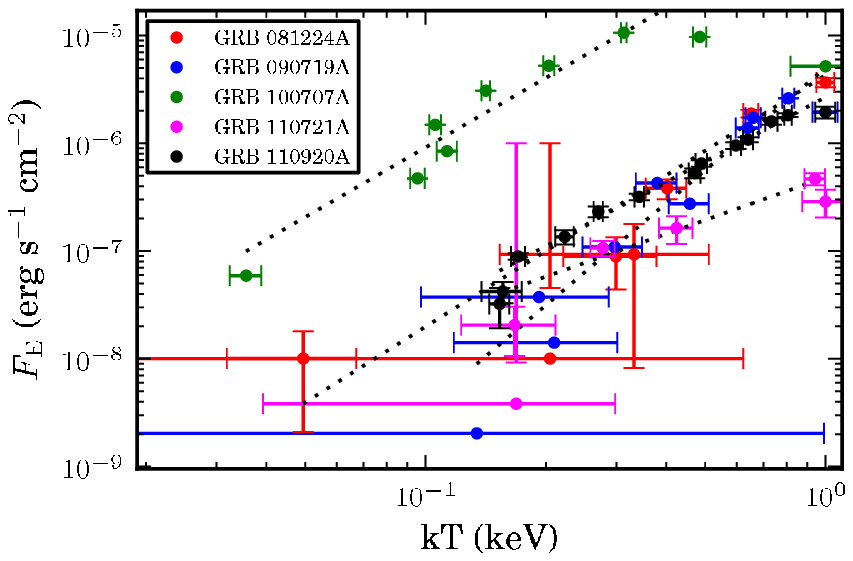}
    }    \subfigure[]{
      \label{fig:kTcor:b}
      \includegraphics[width=8.5cm,height=6.5cm, angle=0]{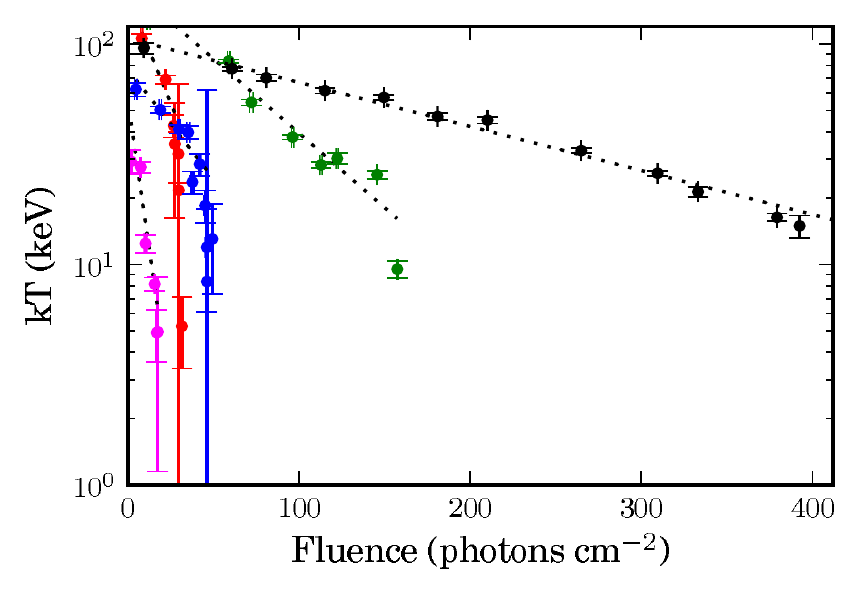}
    }
  \end{center}
  \label{fig:kTcor}

  \caption{The HIC and HFC correlations for the blackbody are separate
    from those derived from the synchrotron component. This adds more
    evidence for the presence of the component. However, the HIC for
    the blackbody is not q=4 as expected unless $\mathcal{R}$ varies
    as is observed.}

\end{figure}

The HIC index for the blackbody component is expected to be 4 provided
N remains a constant in eq. \ref{eq:sbl}; however, nearly all the
blackbodies had a HIC index of $q\sim2$ (Figure
\ref{fig:kTcor:a}). These results confirm those of \citet{Ryde:2001}
and \citet{Ryde:2005} who fit BATSE spectra with a combination of a
blackbody and a power-law to account for the non-thermal
component. There is no physical reason for $\mathcal{R}$ to remain
constant and therefore deviations of $q$ from 4 for the blackbody
component are physically allowed.


Another interesting quantity that can be obtained from the blackbody
is the HFC. All GRBs in our blackbody subset had blackbodies
consistent with the HFC (Figure \ref{fig:kTcor:b}). The decay
constants were all of similar value. \citet{Crider:1999} noted that
similar values of $\Phi_0$ for non-thermal components arise as a
consequence of a narrow parent distribution. A deeper investigation of
a larger sample is required to assess if the same is true for the
blackbody components.

\begin{figure}

  \centering

  \subfigure[]{
    \label{fig:ktEvo:a}
    \includegraphics[scale=.9, angle=0]{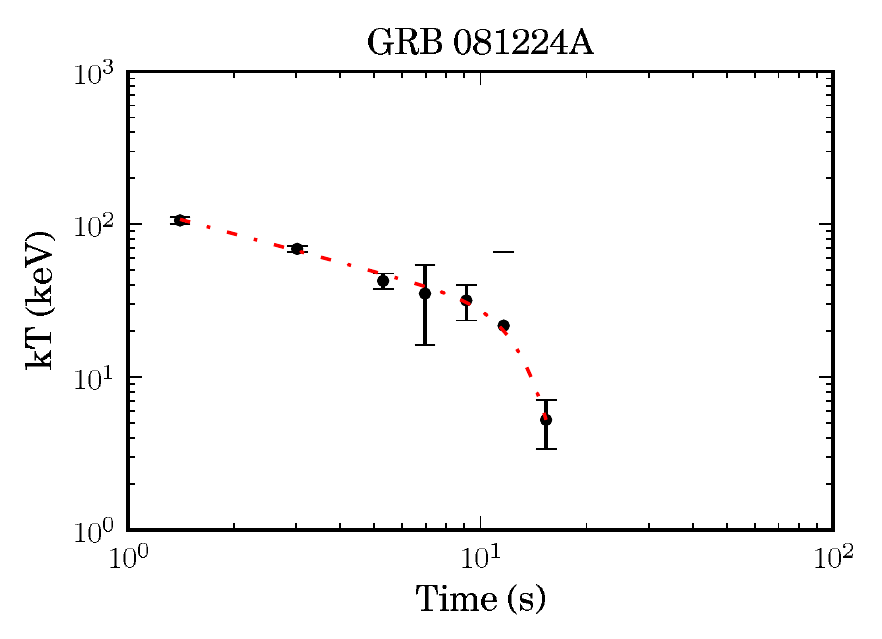}}\subfigure[]{
    \label{fig:ktEvo:b}
    \includegraphics[scale=.9, angle=0]{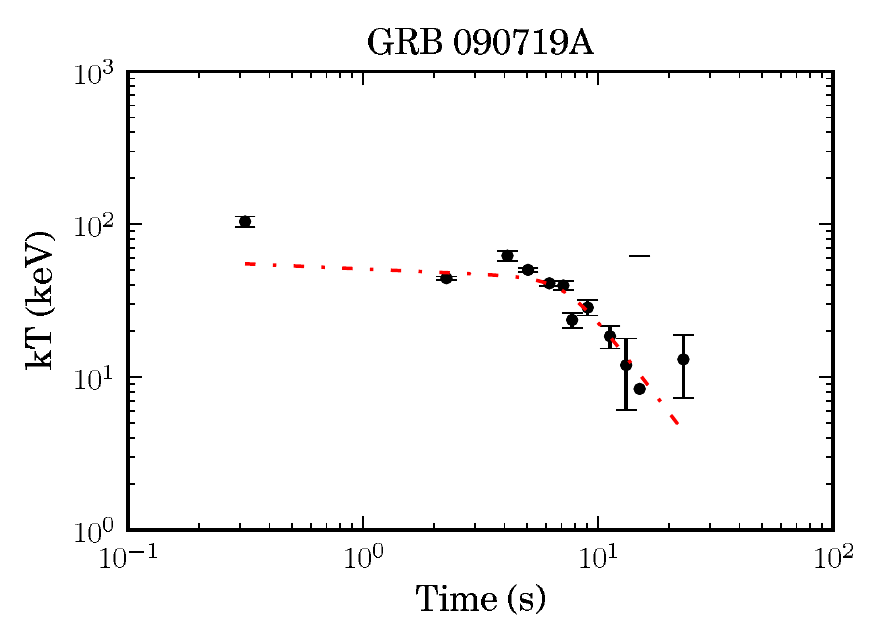}}
\subfigure[]{
    \label{fig:ktEvo:c}
    \includegraphics[scale=.9, angle=0]{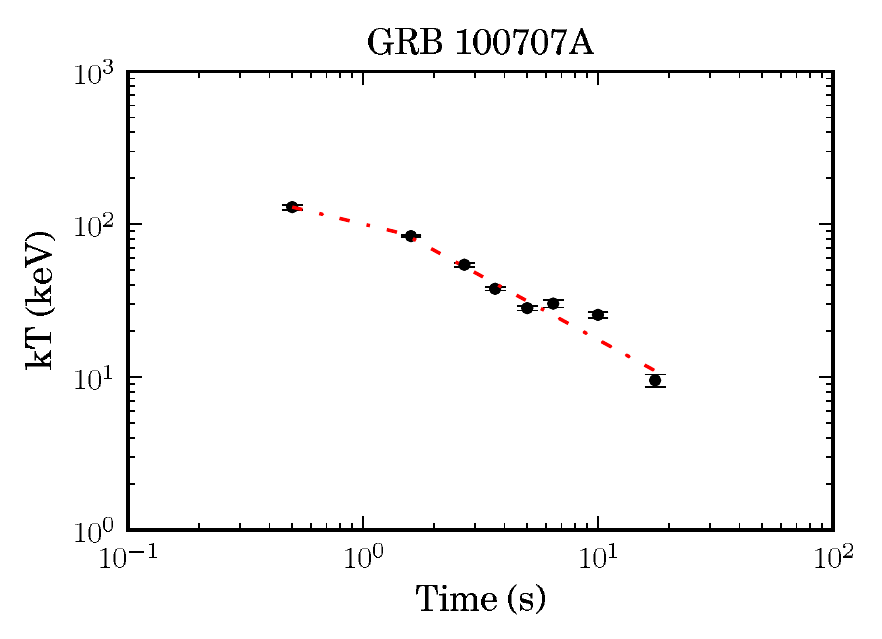}}\subfigure[]{
    \label{fig:ktEvo:d}
    \includegraphics[scale=.9, angle=0]{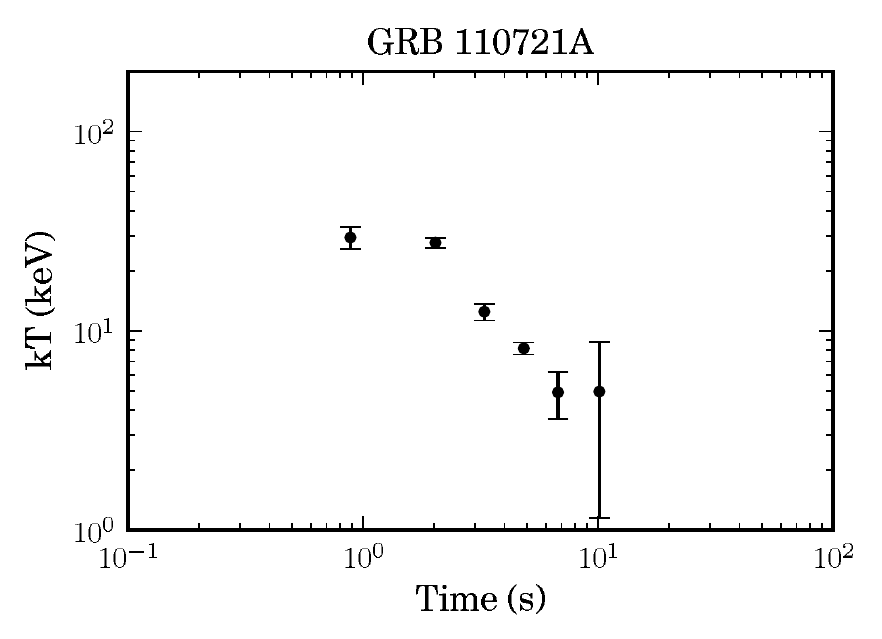}}
\subfigure[]{
    \label{fig:ktEvo:e}
    \includegraphics[scale=.9, angle=0]{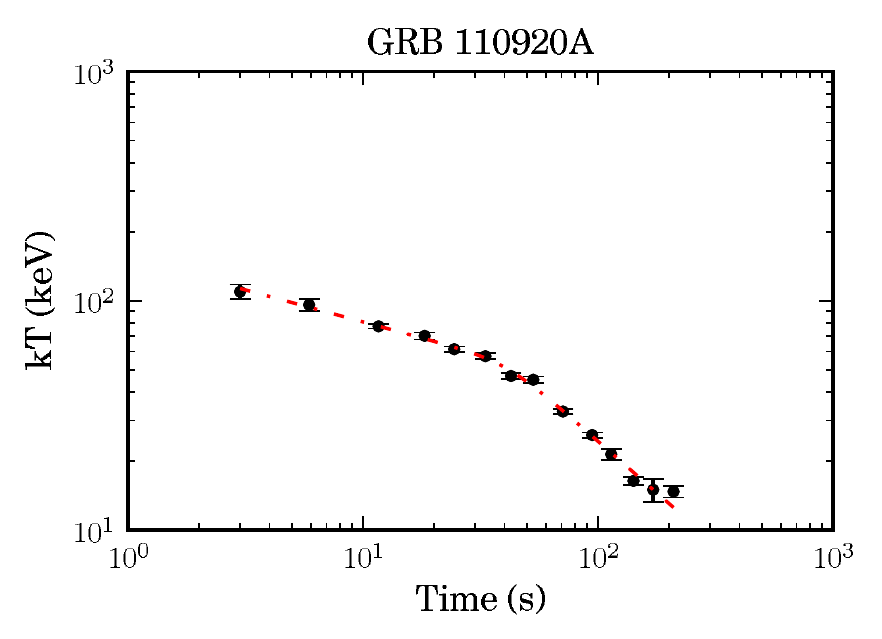}
}

\caption{The time evolution of kT for four of the GRBs in our
  sample. GRB110721A is shown without a fit because the coarse time
  binning used did not allow for constraining the fit
  parameters. However, in \citet{Axelsson:2012}, the evolution is
  shown to follow a broken power-law.}

   \label{fig:kTEvo}

 \end{figure}

The temporal evolution of kT for the blackbody of each burst appears
to follow a broken power-law (Figure \ref{fig:kTEvo}). The evolution
is fit with the function derived in \citet{Ryde:2004} where we fixed
the curvature parameter, $\delta$, to 0.15. The coarse time binning
derived from Bayesian blocks does not allow for the decay indices to
be constrained for all the bursts but a small subset are close to
$-$2/3, as expected (see Table \ref{tab:bbEvo}). The temporal decay
of the blackbody is different than the power-law decay of $E_*$,
indicating a different emission component.

\begin{figure}

  \centering

\subfigure[]{
    \label{fig:scR:a}
    \includegraphics[scale=.9, angle=0]{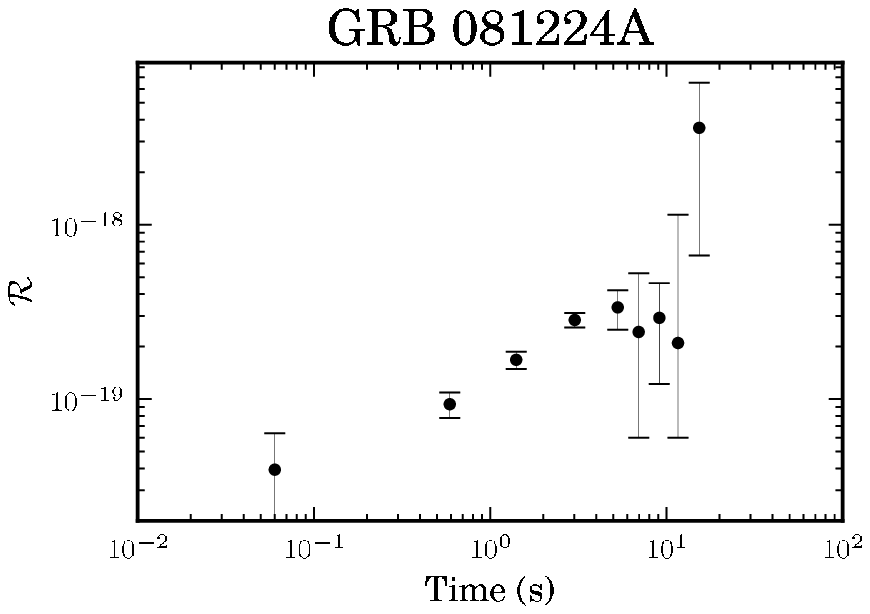}}\subfigure[]{
    \label{fig:scR:b}
    \includegraphics[scale=.9, angle=0]{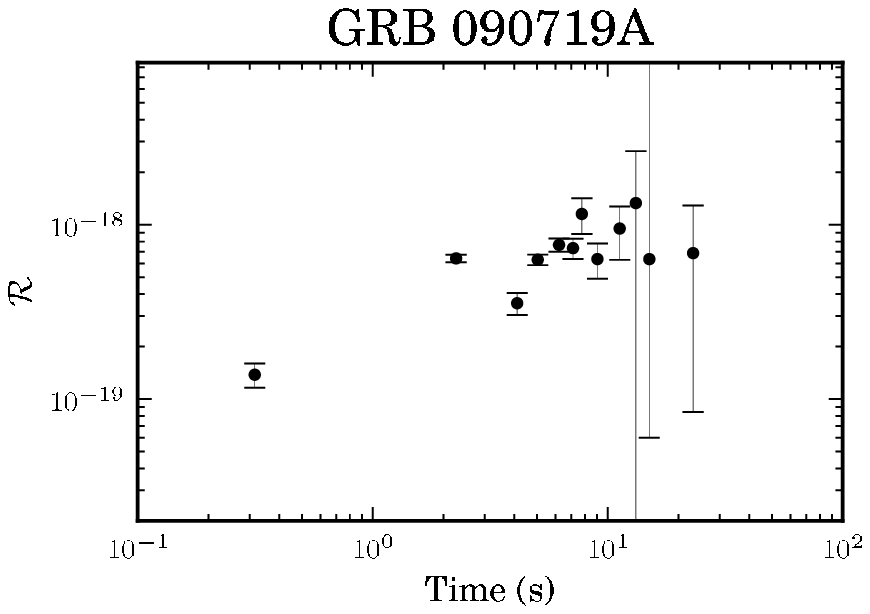}}
\subfigure[]{
    \label{fig:scR:c}
    \includegraphics[scale=.9, angle=0]{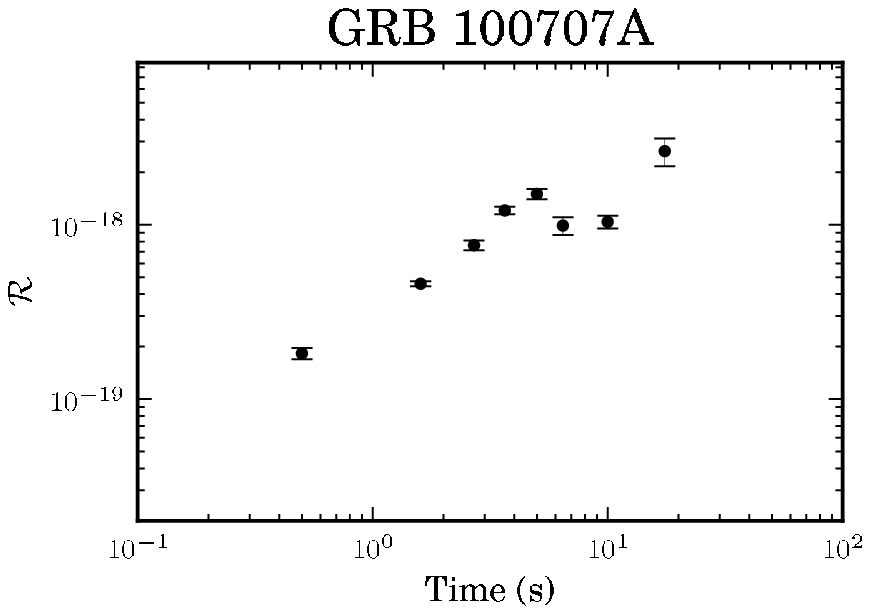}}\subfigure[]{
    \label{fig:scR:d}
    \includegraphics[scale=.9, angle=0]{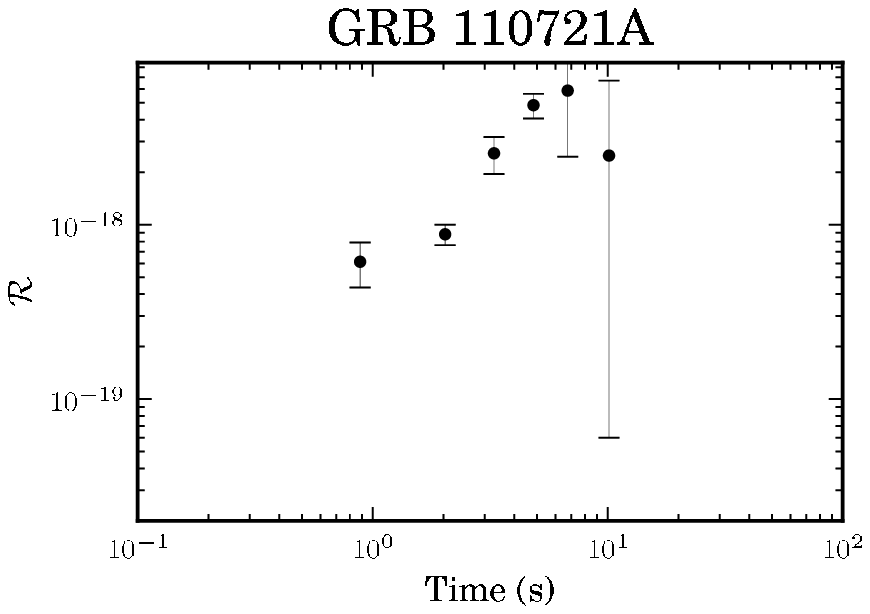}}
\subfigure[]{
    \label{fig:scR:e}
    \includegraphics[scale=.9, angle=0]{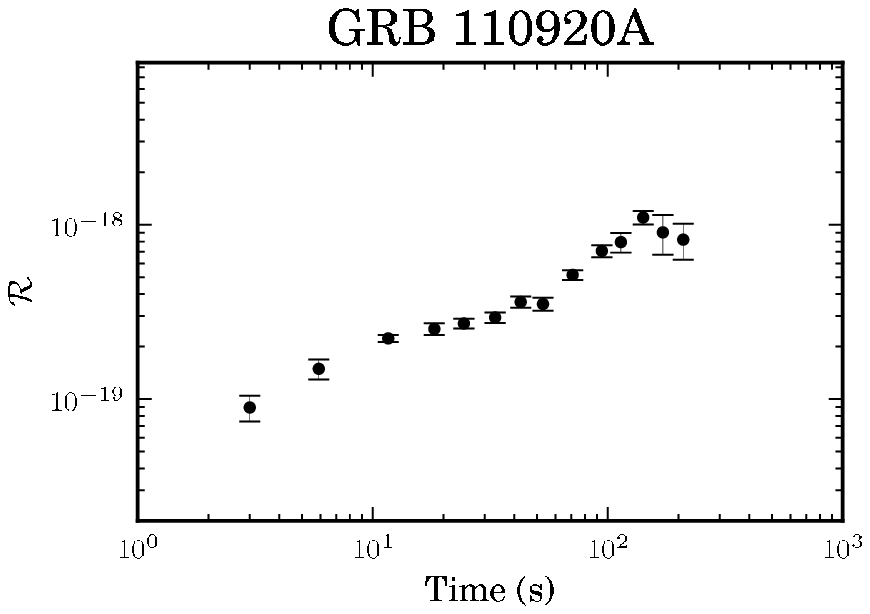}
}

\caption{The evolution of the $\mathcal{R}$ parameter is increases
  with time and shows no relation to the photon flux of the blackbody
  component.}
   \label{fig:scR}

 \end{figure}

The $\mathcal{R}$ parameter was observed to increase with time for all
the GRBs ( Figure \ref{fig:scR}). There are breaks and
plateau in the trends that do not seem to correlate with
the breaks observed in the evolution of kT or with the flux history of
the blackbody. \citet{Ryde:2009} found that the evolution of
$\mathcal{R}$ can be quite complex but mostly follows an increasing
power-law that seems independent of the flux history even for very
complex GRBs. For those complex GRB lightcurves, it was found that
analyzing different intervals of the overlapping pulses yielded an HIC
index for the blackbody of $q\sim 4$ for each interval. In those
intervals, $\mathcal{R}$ was approximately constant indicating the
emission size of the photosphere was constant. Owing to the small
number of time bins, it is difficult to quantify the evolution of
$\mathcal{R}$ for the single pulse GRBs of this study with the coarse
time binning used, but the fact that the HIC index for the blackbodies
differs from $q\sim 4$ and that $\mathcal{R}$ increases indicates that
the evolution of the photosphere is very complex.

\section{Discussion }
\label{sec:discussion}
\subsection{ Importance of Fitting with Physical Models}

By using physical synchrotron emissivities in analysis of 
GRB data, we have
shown that the {\it Fermi} data are consistent with synchrotron
emission from electrons that have not cooled (i.e, slow-cooling
spectra) and are inconsistent with synchrotron emission from electrons
that are cooling (fast-cooling).  The method leads to some interesting
conclusions for empirical modeling. 
There is a positive correlation between hard $\alpha$ and the
inability of model to fit the data, but the low-energy
index of synchrotron seems to be clustered around $-$0.8 rather than near the
asymptotic $-$2/3 found in \citet{Preece:1998}. Not only do the fits
with Band $\alpha$ near $-$0.8 lead to better fits with
synchrotron, but simulated synchrotron spectra are best fit with a
Band function having $\alpha\approx -0.8$.

Previously, GRB spectra have been successfully fitted with a thermal +
non-thermal model by using a blackbody function combined with a Band
function. A thermal spectral component has indeed been shown to be
significant in several cases, foremost in GRB 100724B and in GRB
110721A \citep{Guiriec:2011,Axelsson:2012}. The Band function in these
fits is, however, not based on any physical arguments but is merely an
empirical function that has a convenient parameterization. A general
problem that arises in this type of fitting is that Band $\alpha$ and
the strength of the blackbody component give fits with degenerate
parameters. When the data are fit by a Band function alone, even if an
additional component really exists in the data at lower energies, it
may not be identified because the Band $\alpha$ can accommodate the
additional low energy flux by changing its slope.

The slow-cooling synchrotron model we use here is more restrictive
compared to the Band function. In particular, a limit to the low
energy slope and the curvature of the spectrum are predicted by the
model.  We find that the spectrum below the synchrotron $\vFv$ peak is
not always satisfactorily fit using just the synchrotron model. Except
for extreme cases such as fast and marginally fast-cooling, which
affect the width of the peak as much as the low-energy index, we find
that the low-energy photon spectrum is actually well modeled with a
slope equal to the low-energy slope of the single particle synchrotron
emissivity. This is only possible with very low-radiative efficiency
if the standard GRB acceleration model described in Section 2.1 is
considered.

In many of our fits an additional component is suggested by the
residuals, and the simulations show that this additional component is
statistically significant. Additional components can also be favored
in GRB spectra fit with the Band function, but we find that the
significance of the additional component can be greater when using
physical synchrotron emission fits than when using Band fits. Because
the Band function can accommodate the extra emission using a suitable
power-law index $\alpha$, but the synchrotron function is more
restrictive, an additional component may be more significantly
required when using synchrotron emission for a prescribed electron
distribution.

Another point in Figure \ref{fig:fluxComp} is that when using the
synchrotron model, the temporal evolution of the blackbody flux
exhibits well-defined pulses and a spectral evolution that is clearly
separated from the non-thermal emission. This is in contrast to the
less smooth blackbody flux variations when using the Band function as
the non-thermal process. This fact again reflects that the Band
function is less restrictive than the synchrotron function and thereby
gives rise to further scatter in the derived fluxes in the light
curves. These results suggest that:
\begin{enumerate}[(i)]
 \item the synchrotron function is a good physical model to use;
 \item the thermal component does exist; and
 \item multi-component fitting with the Band function can be misleading.
\end{enumerate}

\subsection{Alleviating Problems with Synchrotron Models}
The fact that the non-thermal spectra seem to be consistent with
slow-cooled synchrotron rather than the fast-cooled synchrotron regime
places strong constraints on the emission model of GRBs.  The
low-energy spectral index of 11 bright BATSE GRBs fall between the
cooled and uncooled limits \citep{Cohen:1997}.
\citet{Ghisellini:2000} showed that it was difficult to reconcile the
implied fast-cooling from a comparison of cooling and dynamical
timescales with the many GRB spectra that require a slow-cooling
electron distribution, leading to spectral problems for the internal
shock model. In Appendix A, we show that a weak-cooled system requires
$\lesssim 100$ G fields for typical bright GRBs detected with {\it
  Fermi}, rather than 100 kG fields, with typical electron Lorentz
factors $\gamma^\prime \approx 10,000$ rather than $300$.

In our simple strong-field synchrotron model, we can neglect the
effects of Compton cooling, which can significantly alter the value of
the low-energy slope in certain parameter regimes
\citep{Daigne:2011}. This could make some spectra less consistent with
fast-cooling, but requires further study.  Klein-Nishina effects on
Compton cooling were not considered, but in the absence of extra
spectral components, either from SSC, hadronic emissions, or external
Compton processes, our synchrotron study is consistent. The need for a
slow-cooling scenario, or marginally slow-cooling system in order to
have reasonable radiative efficiency, is obtained in external shock
model calculations by choosing the $\epsilon_B$ parameter $\approx
10^{-3}$ -- 10$^{-4}$ \citep{Chiang:1999}.


The fast-cooling internal-shock scenario cannot be reconciled with our
observations. Additionally, \citet{Iyyani:2013} found that for GRB
110721A, the standard slow-cooling synchrotron scenario from impulsive
energy input such as internal shocks places the non-thermal emission
region below the photosphere. This may be understood if the electrons
are highly radiative, yet without displaying a cooling spectrum.


Models with ongoing acceleration via first-order and second-order
Fermi acceleration \citep{Waxman:1995,Dermer:2001}, or magnetic
reconnection and turbulence models, including the ICMART model
\citep{Zhang:2011}, have the ability to balance synchrotron cooling
with stochastic heating, or to have multiple acceleration events,
which keep $\gamma_{\rm cool}$ above $\gammaMin$, in which case the
spectrum would resemble a slow-cooled synchrotron spectrum.
Magnetized jet or subjet models \citep{Lazar:2009}
can extend the non-thermal emission site far above the photosphere, and 
relativistic MHD turbulence provides an alternative second-order mechanism
\citep[for a recent review of astrophysical turbulence, see][]{Lyutikov:2013}.
In such a scenario, the electrons cool by synchrotron, but are at the
same time subject to ongoing acceleration, contrary to the low implied
value of the cooling frequency. Slow-cooling or fast-heating scenarios
explain the data much better than a fast-cooling internal-shock model,
though the latter is more radiatively efficient.

\subsection{Conclusion}
We have demonstrated that for a set of {\it Fermi} GRBs we can fit a
physical, slow-cooling synchrotron model directly to the data. Most of
the fitted spectra also require a weaker blackbody component with a
temperature that places its peak below the synchrotron $\vFv$
peak. The temporal evolution of both radiative components shows how
GRB jet properties change, and are free of some of the assumptions
required when fitting GRB spectra with empirical functions. In our
model, a disordered magnetic field is assumed, which could be shown to
be invalid from X-ray and $\gamma$-ray polarization observations,
which are yet inconclusive.  Several parameters in our model cannot be
separately constrained by the fits, namely $\gamth$, $\gammaMin$, and
B, so we focus on a highly magnetized scenario where the self-Compton
component can be neglected.

We find that the energy flux varies as the peak photon energy $E_{p}$
of the peak of the $\nu F_\nu$ spectrum according to $E_{p}^q$, with
$1.1 \lesssim q \lesssim 2.4$.  The dependence of $E_{p}$ is found to
follow the exponential-decay behavior with accumulated fluence
$\Phi(t)$ given by eq.\ (\ref{eq:hfc}), with decay constant
$\Phi_0\approx 100$ -- $700$ [photons cm$^{-2}$]. (see Fig.\
\ref{fig:Epcor}).  For the GRBs where both synchrotron and blackbody
components can be resolved, we find that their parameters follow a
separate temporal behavior.

The temporally evolving spectra were examined in terms of fast-cooling
and slow-cooling electron distributions, considering parameters for a
highly magnetized GRB jet.  The temporal evolution of both synchrotron
and blackbody parameters imply that in the GRBs studied, a photosphere
is formed below a non-thermal emitting region found at a radius
corresponding to the characteristic internal shock scenario. The
electrons in the non-thermal emitting region must undergo continuous
acceleration to produce an apparently slow-cooling synchrotron
spectrum, which can be provided by magnetic reconnection events or
second-order stochastic gyroresonant acceleration with MHD turbulence
downstream of the forward and reverse shocks formed in shell
collisions.  If, on the other hand, the jet fluid is not strongly
magnetized, then it will be radiatively inefficient and have a strong
inverse Compton component.  The use of physical models provides
stronger constraints on jet model parameters, and in future studies we
can relax choices of electron Lorentz factors and magnetic fields by
considering leptonic Compton cascading, and ultra-high energy cosmic
rays.







\appendix

\label{sec:appendix}

\section{Synchrotron-Shell-Model Constraints}
Broad ranges of parameter values are possible in a GRB colliding shell
model. Here we justify the values used to fit the {\it Fermi} GBM and
LAT GRBs, assuming that the bright keV -- MeV emission of the GRBs in
our sample is primarily nonthermal synchrotron radiation emitted by
nonthermal electrons with an isotropic pitch-angle distribution that
radiate in a spherical shell expanding at relativistic speeds, within
which is entrained randomly directed magnetic field on coherence
length scales small in comparison with the shell volume. For
additional considerations about synchrotron models, see
\citet{2013ApJ...769...69B}.

The constraints that we consider are (1) particle and magnetic-field
energetics; (2) a negligible synchrotron self-Compton (SSC) component
so that we can neglect any high-energy $\gamma$-rays that could be
absorbed through $\gamma\gamma$ pair production and make additional
radiation at energies where the data are fit; (3) small synchrotron
self-absorption; and (4) minimum bulk Lorentz factor $\Gamma_{min}$ to
avoid strong $\gamma\gamma$ opacity. We also examine (5) the criterion
for being in the strong cooling regime.  To suppress SSC, we focus on
magnetically dominated models, which are also required in some
theories of GRBs to trigger magnetic reconnection events and produce
the prompt GRB emission through synchrotron emission
\citep[e.g.,][]{2009JPhCS.189a2018G,Zhang:2011}. Magnetically
dominated GRB synchrotron models are also required for efficient
acceleration of ultra-high energy cosmic rays
\citep{2010OAJ.....3..150R}.

Our fiducial parameters are: characteristic electron Lorentz factor
$\gamma^\prime = 10^3 \gamma^\prime_3$; bulk Lorentz factor $\Gamma =
300\Gamma_{300}$, and fluid magnetic field $B^\prime = 10^5
B^\prime_5$ G. Radiation with characteristic $\nu F_\nu$ peak
frequency $\nu_{obs} = m_ec^2 \epsilon /h(1+z)$ is observed during the
prompt phase of the GRB. If nonthermal lepton synchrotron radiation,
then $\epsilon \cong 3\Gamma B^\prime \gamma^{\prime 2}/2B_{cr}$, and
$z$ is the source redshift, so $B_5^\prime \cong \epsilon/\Gamma_{300}
\gamma_3^{\prime 2}$.

\subsection{Energetics}
The electron energy content ${\cal E}_e^{(\prime)}$ in the source
(comoving) frame is given by ${\cal E}_e^\prime = {\cal E}_e/\Gamma =
N_{e0}\gamma^\prime m_ec^2$, where $N_{e0}$ is the number of
electrons, so that
\begin{equation}
{\cal E}_e = {{\cal E}_{par}\over 1+\zeta}
 = 
{6\pi m_ec L_{syn}\over \sigma_{\rm T} B^{\prime 2} \gamma^\prime \Gamma^2}
=
\,{27 \pi m_ec L_{syn}\over 2  \sigma_{\rm T} B_{cr}^2 \epsilon^2}
\,\Gamma\gamma^{\prime 3}
 \cong 
10^{45} \, {L_{51} \Gamma_3 \gamma_3^{\prime 3}\over \epsilon^{2}}\;\;{\rm erg}\;, 
\label{Epar}
\end{equation}
where the total particle energy is denoted ${\cal E}_{par} $, and
$\zeta$ represents the additional energy in hadrons. Here the
synchrotron luminosity $L_{syn}=10^{51}L_{51}$ erg s$^{-1}$ is derived
from the synchrotron electron energy-loss rate formula, using
$L^\prime_{syn} = c \sigma_{\rm T} B^{\prime 2}\gamma^{\prime
  2}N_{e0}/6\pi$.

The magnetic-field energy density ${\cal E}_B = \Gamma {\cal
  E}_B^\prime = \Gamma 4\pi r^2 \Delta r^\prime (B^{\prime 2}/8\pi)$.
The shell width $\Delta r^\prime = k r/\Gamma$, with $k$ a factor of
order unity \citep[for details, see][]{2013ApJ...769...69B}, using the
relations $\Delta r^\prime \cong \Gamma c t_{var}$ and $r \cong
\Gamma^2 c t_{var}$, where $t_{var}$ is the measured variability time
scale in the source frame. Thus the isotropic magnetic-field energy
\begin{equation}
{\cal E}_{B} = {2k \Gamma^4\over 9}\, {c^3 t_{var}^3 B_{cr}^2 \epsilon^2\over \gamma^{\prime 4}} \cong 
10^{56}  k({\Gamma_{300}\over \gamma_3^{\prime}})^4 t_{var}({\rm s})^3 \epsilon^{2}\;\;{\rm erg}.
\label{EB}
\end{equation}
The absolute magnetic field energy ${\cal E}_{B,abs} \cong
(\theta_j^2/2){\cal E}_{B}$ for this system greatly out of
equipartition can be reduced to acceptable values (i.e., ${\cal
  E}_{abs}\ll 10^{54}$ erg) with a sufficiently small jet opening
angle $\theta_j$ between $\approx 0.01$ and $0.1$.

\subsection{SSC Component}
The ratio of the SSC and synchrotron luminosities is related to the
ratio of the synchrotron and magnetic field energy densities through
the relation $L_{SSC}/L_{syn} \lesssim
u^\prime_{syn}/u^\prime_{B^\prime}$, with the inequality arising from
the neglect of Klein-Nishina effects on the SSC emission.  Because
$u^\prime_{syn} \cong L^\prime_{syn}/4\pi r^2 c$, we have
\begin{equation}
{L_{SSC}\over L_{syn}} \approx {2 L_{syn}\over c^3 \Gamma^6 t_{var}^2 B^{\prime 2}} \cong {10^{-5}  L_{52}\over \Gamma_{300}^6 t_{var}^2({\rm s}) B^{\prime 2}_5 } \;,
\label{SSCsynratio}
\end{equation}
and so can be safely neglected here.

\subsection{Synchrotron Self-Absorption}
For a log-parabolic description of the $\gamma^{\prime 2} N^\prime
(\gamma_p)$ electron distribution, the SSA opacity in the
$\delta$-function approximation is given by
\begin{equation}
  \tau_{\epsilon^\prime} = 2\kappa_{\epsilon^\prime} \Delta r^\prime \cong {\pi \over 9}\,{{\cal E}_e^\prime \Delta r^\prime\over m_ec^2 I(b) V_b^\prime \gamma_p^{\prime 4}}\, {\lambda_{\rm C} r_e\over \epsilon^\prime} (2+ b\log x)\,x^{-(4+b\log x)}
  \equiv \tau_0 (2+ b\log x)\,x^{-(4+b\log x)}\;
\label{tauep}
\end{equation}
\citep{2009herb.book.....D,2013arXiv1304.6680D}, where
$\kappa_{\epsilon^\prime}$ is the SSA absorption coefficient (units of
inverse length), $x \equiv
\sqrt{\ep/2\varepsilon_B^\prime}/\g_p^\prime$, $\e \cong \Gamma\ep$,
shell volume $V_b^\prime = 4\pi r^2 \Delta r^\prime$, and $I(b) =
\sqrt{\pi \ln 10/b}$ normalizes the electron spectrum depending on the
value of the log-parabola width parameter $b$.  Using eq.\ (A2), we
obtain
\begin{equation}
\tau_{0} \cong {\pi \over 6\e }\,{ {\lambda_{\rm C} r_e L_{syn} \over c^3 \sigma_{\rm T} B^{\prime 2} t_{var}^2 \Gamma^5 \gamma^{\prime 5} I(b)}}
\approx {10^{-16} \over \e^3 }\,{ { L_{51} \over  t_{var}^2({\rm s}) \Gamma_{300}^3 \gamma_3^\prime I(b)}}\;, 
\label{tau0}
\end{equation}
using the relation $\epsilon \cong \Gamma_{300} B^\prime_5
\gamma^{\prime 2}_3$ characterizing the condition that $x \approx 1$.
Thus SSA is utterly negligible at $x \gtrsim 0.1$, where the question
of SSA opacity is most important, noting from eq.\ (A5) that the
opacity can grow as fast as $x^{-4}$ at $x\approx 0.1$, when
$b\lesssim 1$.

\subsection{$\gamma$-$\gamma$ Opacity}
A $\gamma$-ray with energy $\epsilon_\gamma =1.96 \times 10^5
E_\gamma$(GeV) is subject to absorption through the pair-production
process $\gamma\gamma\rightarrow e^\pm$ when passing through a target
radiation field.  The minimum bulk Lorentz factor $\Gamma_{min}$ that
gives unity optical depth for absorption by target synchrotron photons
is estimated within $\approx 10$\% accuracy by the expression
\begin{equation}
\Gamma \geq \Gamma_{min} = \left[ { \sigma_{\rm T} \hat\epsilon L(\hat \epsilon ) \epsilon_{\gamma}
\over 16 \pi m_ec^4 t_{var} }\right ]^{1/6}\;,\; \hat \epsilon \cong 2\Gamma^2/\epsilon_\gamma .
\label{tauep}
\end{equation}
Taking $\e L(\e) \cong 10^{51}L_{51}/\ln (100)$ erg s$^{-1}$, i.e., a
flat $\nu F_\nu$ spectrum over 2 decades in frequency, then the
minimum bulk Lorentz factor $\Gamma_{min} \approx 300 \,[L_{51}
E_\gamma$(100~GeV)$/t_{var}({\rm s})]^{1/6}$.  For GRB synchrotron
radiation emitted in the $0.1 \lesssim \epsilon \lesssim 10$ range,
$\gamma$-rays with energies between $\approx (0.01$ --
1)$\Gamma_{300}^2$ TeV are subject to $\gamma\gamma$ opacity.
Provided that the energy radiated at 100 GeV and TeV energies is much
smaller that the total GRB photon energy, opacity effects and
cascading can be neglected.

\subsection{Cooling Regime}

The minimum and cooling frequencies in a colliding shell are derived
in the same way as the case of a blast wave decelerating by sweeping
up external medium material at a shock \citep{Sari:1998}, recognizing
that the relative Lorentz factor between two shells is more likely to
be $\Gamma_{rel}\sim 10$, compared to the external shock Lorentz
factor $\Gamma \sim 300$. The system is in the slow cooling regime
when the cooling Lorentz factor
\begin{equation}
\gamma_c^\prime \cong {6\pi m_e c\over \sigma_{\rm T} B^{\prime 2}\Gamma t_{var} }
\gtrsim
\gamma^\prime_{min} \cong \epsilon_e {m_p\over m_e} f(p) \Gamma_{rel}\;,
\label{gammaprimemin}
\end{equation}
where $\gamma^\prime_{min}$ is the minimum electron Lorentz factor,
$p$ is the injection number index of relativistic electrons,
$\epsilon_e$ is the fraction of energy dissipated at the shock that
goes into nonthermal electrons, and the factor $f(p) = (p-2)/(p-1)$
normalizes the number and energy of the energized electrons.  Solving
gives
\begin{equation}
B^{\prime}
\lesssim 
\sqrt{ 6\pi m_e c (m_e/m_p)\over \sigma_{\rm T} \Gamma t_{var}\epsilon_e f(p)\Gamma_{rel} }
\approx {120 {\rm ~G}\over \sqrt{\Gamma_{300} (\epsilon_e/0.1) t_{var}({\rm s}) f(p)\Gamma_{rel}}}\;.
\label{gammaprimemin}
\end{equation}
A system with $\sim 100$ kG fields is always in the fast cooling
regime according to this criterion.

\acknowledgements

We would like to thank Fr\'ed\'eric Daigne and Bing Zhang for helpful
discussions concerning the nature of the non-thermal emission, and the
referee for a constructive report.  We would also like to thank Paz
Beniamini and Tsvi Piran for extremely helpful comments clarifying the
limitations of this model, now addressed in the Appendix. The work of
C.D.D. is supported by the Office of Naval Research.

The Fermi GBM collaboration acknowledges support for GBM development,
operations and data analysis from NASA in the US and BMWi/DLR in
Germany.

The \textit{Fermi} LAT Collaboration acknowledges generous ongoing
support from a number of agencies and institutes that have supported
both the development and the operation of the LAT as well as
scientific data analysis.  These include the National Aeronautics and
Space Administration and the Department of Energy in the United
States, the Commissariat \`a l'Energie Atomique and the Centre
National de la Recherche Scientifique / Institut National de Physique
Nucl\'eaire et de Physique des Particules in France, the Agenzia
Spaziale Italiana and the Istituto Nazionale di Fisica Nucleare in
Italy, the Ministry of Education, Culture, Sports, Science and
Technology (MEXT), High Energy Accelerator Research Organization (KEK)
and Japan Aerospace Exploration Agency (JAXA) in Japan, and the
K.~A.~Wallenberg Foundation, the Swedish Research Council and the
Swedish National Space Board in Sweden.
 
Additional support for science analysis during the operations phase is
gratefully acknowledged from the Istituto Nazionale di Astrofisica in
Italy and the Centre National d'\'Etudes Spatiales in France.


\def\mn{M.N.R.A.S.}  \def\aas{{Astron. Astrophys.}}

\def\aassupp{{Astron. Astrophys. Supp.}}  \def\apss{{Astr. Space

    Sci.}}  \def\apj{ApJ} \def\apjl{ApJ} \def\nat{Nature}

\def\aaps{{Astron. \& Astr. Supp.}}  \def\aa{{A\&A}} \def\apjs{{ApJS}}

\def\sp{{Solar Phys.}}  \def\jgr{{J. Geophys. Res.}}

\def\grl{{Geophys. Res. Lett.}}  \def\jphysb{{J. Phys. B}}

\def\ssr{{Space Science Rev.}}

\def\araa{{Ann. Rev. Astron. Astrophys.}}  \def\nature{{Nature}}

\def\asr{{Adv. Space. Res.}}  \def\prc{{Phys. Rev. C}}

\def\prd{{Phys. Rev. D}} \def\pr{{Phys. Rev.}}

 \def\prl{{Phys. Rev. Lett.}}

\newpage


%

%

%

%


\clearpage

\begin{table}

\centering

\begin{tabular}{c | c c c c}

GRB & Peak Flux (p/s/cm$^2$) & Duration Analyzed (s) & Blackbody Component & LLE data\\ 

\hline \hline

GRB 081110A & 20.88  &  4.61 &   \text{\sffamily X}         & \text{\sffamily X}\\ 

GRB 081224A & 17.11  & 18.36  & $\checkmark$ & $\checkmark$\\

GRB 090719A & 26.52   & 30.09  & $\checkmark$ & \text{\sffamily X}\\

GRB 090809B & 18.36  & 14.64  & \text{\sffamily X} & \text{\sffamily X}\\

GRB 100707A & 18.77  & 22.39  & $\checkmark$ & \text{\sffamily X}\\

GRB 110407A & 15.6  & 20.48  & \text{\sffamily X} & \text{\sffamily X}\\

GRB 110721A & 29.82   & 12.7  & $\checkmark$ & $\checkmark$\\

GRB 110920A & 8.08   & 238.29 & $\checkmark$ & \text{\sffamily X} \\

\end{tabular}

\caption{The GRBS in our sample. The peak fluxes were taken from the brightest bin of each GRB with a duration determined by Bayesian blocks.}

\label{tab:grbs}

\end{table}

\begin{table}

\centering

\begin{tabular}{c | c c c c}

Time Bin & Band C-Stat & Band+BB C-Stat & Synchrotron C-stat & Synchrotron+BB C-stat \\ 

\hline \hline

-0.2-0.2 & 453 & 450 & 485 & 455 \\

0.2-0.8 & 374 & 362 & 695 & 364 \\

0.8-2.4 & 427 & 405 & 2546 & 487 \\

2.4-3.0 & 408 & 400 & 903 & 413 \\

3.0-4.3 & 431 & 415 & 1214 & 430 \\

4.3-5.7 & 397 & 363 & 791 & 399 \\

5.7-7.2 & 411 & 390 & 598 & 422 \\

7.2-12.8 & 488 & 414 & 829 & 447 \\

12.8-22.2 & 558 & 412 & 594 & 423 \\

\end{tabular}

\caption{The time resolved C-Stat values for GRB100707A show that while the Band function and synchrotron models combined with a blackbody function both fit the data well, the non-thermal functions fit the data very differently when not combined with a blackbody. Specifically, where the blackbody is the brightest (Figures \ref{fig:fluxComp:c} and \ref{fig:fluxComp:d} intervals 2, 3, and 4) the Band function alone fits the data acceptably while the synchrotron model alone fits the data poorly. This shows that the flexibility of the Band function can mask the need for the blackbody component. The Band+blackbody fits actually fit the data better when the blackbody is very bright in this case. This is most likely due to the blackbody function (eq. \ref{eq:blackbody}) used is simplified and the actual emission may be broadened due to beaming effects that are only important to the fit when the blackbody is bright and the synchrotron fit is used. The Band function makes up for these effects by having a harder $\alpha$. We tested using an exponentially cutoff power-law combined with the synchrotron model and the fits were as good as those with the Band function. We will examine the use of a more realistic photosphere model in future work.}

\label{tab:grb1c}

\end{table}

\begin{table}

\centering

\begin{tabular}{c | c c c c c c}

Time Bin & Band C-Stat & Band+BB C-Stat & Synchrotron C-stat & Synchrotron+BB C-stat & Fast C-stat & Fast+BB C-stat \\ 

\hline \hline

-0.07-0.08 & 640 & 640 & 673 & 673 & 709 & 709 \\

0.08-0.48 & 690 & 690 & 704 & 704 & 1088 & 1088 \\

0.48-1.28 & 709 & 668 & 688 & 670 & 1654 & 957 \\

1.28-2.78 & 887 & 761 & 838 & 770 & 1646 & 1041 \\

2.78-3.78 & 678 & 642 & 655 & 643 & 797 & 666 \\

3.78-5.88 & 648 & 631 & 677 & 634 & 694 & 660 \\

5.88-7.63 & 729 & 728 & 733 & 721 & 773 & 724 \\

7.63-12.63 & 932 & 693 & 693 & 692 & 756 & 698 \\

\end{tabular}

\caption{The C-stat values for GRB 110721A. The significance of the addition of the blackbody is not as large as with GRB 100707A (Table \ref{tab:grb1c}) due to the weakness of the blackbody component. The fits for fast-cooled synchrotron are included to demonstrate the poor quality fits that are obtained both with fast-cooling synchrotron and fast-cooling synchrotron with a blackbody.}

\label{tab:grb2c}

\end{table}

\begin{table}

\centering

\begin{tabular}{c | c | c c | c}

Time Bin & Band $\alpha$ & Slow-cooled Synchrotron C-stat & Fast-cooled Synchrotron C-stat & $\Delta_{\rm C-stat}$ \\ 

\hline \hline

-5.38-2.82 & -0.9 & 523 & 599 & 76 \\

2.82-3.84 & -0.7 & 507 & 604 & 97 \\

3.84-4.86 & -0.8 & 506 & 596 & 90 \\

4.86-6.91 & -1.0 & 534 & 626 & 92 \\

6.91-9.98 & -1.1 & 591 & 639 & 48 \\

9.98-15.1 & -1.5 & 494 & 494 & 0 \\

\end{tabular}

\caption{For each time bin of GRB 110407A we examine the C-stat value of synchrotron and fast-cooled synchrotron. This GRB did not have a blackbody in its spectrum. While the Band function and slow-cooling synchrotron fit the spectrum well, fast-cooling synchrotron does not fit the spectrum unless Band $\alpha=-1.5$. In this case, the fast-cooled synchrotron peak energy was very unconstrained due to the curvature of the data being narrower than the photon model's curvature.}

\label{tab:fast}

\end{table}

%



\begin{table}
\centering
\begin{tabular}{c | c c c c}
GRB & Flux Index $q$ & $\chi^2_{red}$ & $\Phi_0$ & $\chi^2_{red}$ \\ 
\hline \hline
GRB 081110A & 2.32$\pm$0.4 & 0.6 & 97$\pm$23 & 0.4 \\ 

GRB 081224A & 1.74$\pm$0.1 & 1.5 & 253$\pm$23 & 0.3 \\ 

GRB 090719A & 1.14$\pm$0.07 & 0.98 & 245$\pm$17 & 1.2\\

GRB 09080B & 1.58$\pm$0.05 & 8.0 & 188$\pm$9 & 1.0 \\ 

GRB 100707A & 1.04$\pm$0.02 & 1.2 & 444$\pm$24 & 7.3 \\ 

GRB 110407A & 1.72$\pm$0.20 & 0.5 & 214$\pm$32 & 4.2 \\ 

GRB 110721A & 1.08$\pm$0.03 & 14.4 & 269$\pm$13 & 15.4 \\ 

GRB 110920A & 1.37$\pm$0.06 & 0.5 & 669$\pm$33 & 1.2 \\

\end{tabular}
\caption{Sample correlations for both flux and fluence for the synchrotron component.}
\label{tab:cor}
\end{table}

\begin{table}
\centering
\begin{tabular}{c | c c}

Time Bin & Band-BB $\Delta_{C-stat}$ & Synchrotron-BB $\Delta_{C-stat}$ \\ 
\hline \hline

-0.2-0.2 & 3 & 30 \\ 

0.2-0.8 & 12 & 331 \\ 

0.8-2.4 & 22 & 2059 \\ 

2.4-3.0 & 8 & 490 \\ 

3.0-4.3 & 16 & 784 \\ 

4.3-5.7 & 34 & 392 \\ 

5.7-7.2 & 21 & 176 \\ 

7.2-12.8 & 74 & 382 \\ 

12.8-22.2 & 146 & 171 \\ 

\end{tabular}

\caption{The $\Delta_{C-stat}$ between the Band function and synchrotron model fits with and without the inclusion of a blackbody for GRB 100707A. The blackbody has a significantly larger impact on the fit when included with the synchrotron model.}
\label{tab:grb1dc}
\end{table}

\begin{table}
\centering
\begin{tabular}{c | c c c}

Time Bin & Band-BB $\Delta_{C-stat}$ & Synchrotron-BB $\Delta_{C-stat}$ & Fast-BB $\Delta_{C-stat}$ \\ 
\hline \hline

-0.07-0.08 & 0 & 0 & 0 \\ 

0.08-0.48 & 0 & 0 & 0 \\ 

0.48-1.28 & 41 & 18 & 697 \\ 

1.28-2.78 & 126 & 68 & 605 \\ 

2.78-3.78 & 36 & 12 & 131 \\ 

3.78-5.88 & 17 & 43 & 34 \\ 

5.88-7.63 & 1 & 12 & 49 \\ 

7.63-12.63 & 239 & 1 & 58 \\

\end{tabular}

\caption{ The $\Delta_{C-stat}$ values for GRB 110721A tell a different story than GRB 100707A (Table \ref{tab:grb1dc}), though both GRBs show a significant improvement in the fit when a blackbody is included. Even thought the fast-cooled fits showed extreme improvement with the inclusion of a blackbody, the fits are still poor compared with the slow-cooled model (See Table \ref{tab:grb2c}).}
\label{tab:grb2dc}
\end{table}

\begin{table}
\centering
\begin{tabular}{c| c c c c}
GRB & Flux Index & $\chi^2_{red}$ & $\Phi_0$ & $\chi^2_{red}$ \\
\hline \hline
GRB 081224A & 2.3 $\pm$ 0.3 & 1.4 & 121 $\pm$ 13 & 9 \\ 

GRB 090719A & 2.8 $\pm$ 0.4 & 2.3 & 232 $\pm$ 23 & 3 \\ 

GRB 100707A & 2.2 $\pm$ 0.1 & 17.4 & 319 $\pm$ 8 & 28 \\ 

GRB 110721A & 1.3 $\pm$ 0.2 & 1.8 & 43 $\pm$ 3 & 5 \\ 

GRB 110920A & 2.0 $\pm$ 0.1 & 0.9 & 1147 $\pm$ 21.7 & 4 \\

\end{tabular}
\caption{For the subset of bursts that have a strong blackbody component we compute the flux and fluence correlation for the blackbody.}
\label{tab:bbCor}
\end{table}

\begin{table}
\centering
\begin{tabular}{l| c c c c}
GRB & $F_{bb}/F_{syn}$ & $1^{st}$ Decay Index & $2^{nd}$ Decay Index & $\chi^2_{red}$ \\
\hline \hline 

GRB 081224A & 0.3 & -0.6 $\pm$ 0.07 & -20 $\pm$ 243 & 0.5 \\ 

GRB 090719A & 0.4 & -0.1 $\pm$ 0.05 & -2.0 $\pm$ 0.7 & 11.3 \\ 

GRB 100707A & 0.5 & -0.4 $\pm$ 822749 & -0.8 $\pm$ 0.03 & 22.7 \\ 


GRB 110920A & 0.8 & -0.3 $\pm$ 0.03 & -0.9 $\pm$ 0.04 & 2.0\\

\end{tabular}
\caption{The evolution of the blackbody follows a broken power-law. However, the coarse time bins recovered by the Bayesian blocks algorithm make it difficult to constrain the decay indices.}
\label{tab:bbEvo}
\end{table}

\end{document}